\documentclass[aps,titlepage,12pt]{revtex4}
\usepackage{amsfonts}
\usepackage{amsmath}
\usepackage{graphicx}
\usepackage{dcolumn}
\usepackage{bm}
\usepackage{overpic}
\usepackage{booktabs}
\usepackage{color}
\usepackage[justification=raggedright,font={small,sf}, singlelinecheck=false]{caption}
\usepackage{natbib}

\newcommand{\ddiff}{d_{\rm max} - d_{\rm min}}
\newcommand{\tc}{T_{\rm center}}
\newcommand{\tp}{T_{\rm price}}
\newcommand{\ta}{T_{\rm amenity}}
\newcommand{\dreal}{\bar{d}_{\rm real}}
\newcommand{\drand}{\bar{d}_{\rm random}}
\newcommand{\dgreedy}{\bar{d}_{\rm greedy}}
\newcommand{\dreduce}{\Delta d}

\begin{document}
\title{Guiding Self-Organizing Dynamics of Residential Choice in Cities to Reduce Traffic Congestion and Carbon Emissions}

\author{Yu-Qing Liu$^{1,2,\dagger}$, Chen Zhao$^{2}$\footnote{Email: tczxz007@hebtu.edu.cn}}
\thanks{These authors contributed equally to this work.}
\author{Xiao-Yong Yan$^{3}$, Xiaoyue Hou$^{5}$, Chi Ho Yeung$^{4,}$\footnote{Email: chyeung@eduhk.hk}}
\author{An Zeng$^{5,}$\footnote{Email: anzeng@bnu.edu.cn}}

\affiliation{$^{1}$School of Geographical Sciences, Hebei Normal University, Shijiazhuang, Hebei, P.R. China\\
$^{2}$School of Mathematical Sciences, Hebei Normal University, Shijiazhuang, Hebei, P.R. China\\
$^{3}$School of Systems Science, Beijing Jiaotong University, Beijing, P.R. China\\
$^{4}$Department of Science and Environmental Studies, The Education University of Hong Kong, Hong Kong, P.R. China\\
$^{5}$School of Systems Science, Beijing Normal University, Beijing, P.R. China}

\begin{abstract}
Rapid urbanization and growing vehicle ownership exacerbate traffic congestion and prolong commute times. We examine the self-organizing dynamics of residential choice via a hypothetical home-swapping process to mitigate peak-hour traffic congestion and carbon emissions. Specifically, we analyze over $400,000$ trajectories from $9$ days in a major Chinese city, revealing that actual average commuting distance is approximately three times shorter than under random residential distribution, indicating significant self-organization. Notably, city-wide home swapping reduces commuting distance by $50.4\%$, substantially easing traffic congestion, thereby reducing carbon emissions by $77.3\%$. Even with the consideration of socio-demographic factors and individual needs, the reductions remain significant: $8.1\%$-$10.3\%$ in commuting distance and $27.4\%$-$34.4\%$ in carbon emissions. Considering the potential induction of additional non-commuting trips, the reduction in carbon emissions remains substantial. Given the primacy of distance to the city center, polycentric city layouts can enhance these benefits. For validation, we use another dataset covering China's 28 major cities to confirm these findings. Finally, we introduce a data-driven model to elucidate self-organizing dynamics of residential choice and analyze the feasibility of government coordination. These insights demonstrate that a synergistic alignment of residential choices can leverage individual and city-level benefits, effectively alleviating commuting congestion and associated emissions.

\end{abstract}


\maketitle

\section{Introduction}
With the rapid urbanization, an increasing ratio of the world population resides in urban areas~\cite{sun2020dramatic}. While the room for expansion in urban road infrastructure is small to cope with the increasing population, traffic congestion becomes more severe~\cite{ccolak2016understanding,pi2019visual,zhou2022identifying,dasgupta2021spatiotemporal,wang2023aggravated,manfredi2018mobility,dahlmanns2023optimizing,zhang2019scale}. This does not only reduce the daily comfort of city commuters, but also wastes a lot of resources in terms of time and energy, and induce air and noise pollution, impacting the socio-economy of cities at a holistic level.

Specifically, traffic congestion can be divided into non-recurring and recurring congestion. Non-recurring congestion is caused by unexpected events and is usually unpredictable. Recurring congestion, on the other hand, is caused by periodic traffic flows, such as the commuter flow during morning and evening peak hours. This type of congestion is regular and predictable and may constitute a major part of the urban congestion~\cite{afrin2020survey}. To alleviate recurring traffic congestion, many cities have developed a series of measures. These include advocating the use of public transport, staggered work hours, establishing road tolls in peak hours, restricting vehicle usage based on license plate numbers, etc. In addition, researchers have proposed various methods to ease traffic congestion, including traffic flow prediction ~\cite{ren2014predicting,hou2019adaptive,ma2021short}, control of traffic lights~\cite{cao2016unified,de2020development}, time-based travel arrangements~\cite{xu2018planning,yildirimoglu2021staggered} and dynamic traffic assignment~\cite{aghamohammadi2020dynamic, yeung2013physics, yeung2019coordinating}. However, most methods are remedial without fundamentally tackling the source of traffic congestion.

As we all know, the root cause of traffic congestion during peak hours lies in the need of individuals to travel between home and their workplace or school~\cite{liu2018investigating,NIU2025105583,2017transportreport}. The longer the commuting distance, the more likely private motor vehicles are used, which leads to more vehicles on roads, and hence increased traffic congestion during peak hours~\cite{wheaton2004commuting,zhao2020long}. On the contrary, if the commuting distance is short, people can use non-motor vehicles such as bicycles or even walk to their workplace. Thus, there were studies which examine residences, workplaces, and commuting distances using empirical datasets~\cite{huang2018tracking,plaut2006intra,gerber2020workplace,surprenant2013commuting,guo2021fractal,suzuki2012jobs,zhao2011impact,asmussen2024interplay,hu2017different,ta2017understanding,shen2021job}. Studies show that moving or changing jobs can change the spatial arrangement of residences and workplaces for single-commuter and dual-commuter households, reducing their commuting distance and time~\cite{huang2018tracking,surprenant2013commuting}. Although well-designed surveys can indeed provide statistically valid insights into residential preference through carefully constructed samples~\cite{yan2013diversity}, most existing studies were usually based on surveys from a small number of respondents, and to our knowledge, have never revealed the benefit brought by reallocation of residence at the city level.

In this paper, we examine the impact of a hypothetical city-wide {\bf home-swapping} in reducing traffic congestion and alleviating carbon emissions. We used more than 400,000 real individual mobility trajectories over 9 days from a Chinese city, Shijiazhuang, to infer the home and workplace location as well as household members of each individual, and identify all the commuter path during peak hours using Baidu map~\cite{BaiduPlatform}.
We found that randomly re-shuffling residence locations increases the average commuting distance by $2.83$ times, indicating a strong self-organization in the actual choice of residence location. The self-organization~\cite{bak1988self,haken2006information} in our context refers to the emergent spatial patterns of residential locations that arise from decentralized household decisions made based on their local information, without centralized coordination, such that the average commuting distance is still much smaller than that of the random residential distribution. By strategic home swapping, urban congestion during peak hours is greatly reduced since the average commuting distance is reduced by $50.4\%$, which complies with the notion of a 15-minute city~\cite{abbiasov202415,khavarian202315,papadopoulos2023measuring}.

To make home swapping more realistic, we consider the heterogeneity of households and swap homes only if the three socio-demographic factors including the distance to the city center, housing price and amenity accessibility, and needs of individual household members are preserved for both households. In this case, the average commuting distance can be reduced by $13.4\%$ if only socio-demographic factors are considered. By also considering the needs of individual household members, the average commuting distance can still be reduced by $8.1\%$ to $10.3\%$. In all these cases, owing to the combined effect of shorter commuting distance and reduced traffic congestion, carbon emissions from vehicles are also greatly reduced by $77.3\%$ without constraints and by $27.4\%$ to $34.4\%$ when socio-demographic factors and the needs of individual household members are considered, respectively. Among the three socio-demographic factors, the distance to the city center influences home-swapping most, followed by the amenity accessibility and housing price. Based on our findings, we found that establishing a polycentric, i.e. multi-center, city layout can effectively increase the benefit brought by guiding self-organizing residential choice.

While a great benefit of home-swapping has been shown in the simulation on the dataset of Shijiazhuang, we further verify the universality of our results in cities other than Shijiazhuang and investigate the impact of induced traffic by the reduced traffic congestion. Specifically, we examine the impact of home swapping using another origin-destination dataset covering 28 major cities in China~\cite{zhao2024unravelling}, and the results are consistent with our findings in Shijiazhuang, suggesting universal benefits can be brought by home swapping. Since traffic congestion is extensively reduced by home swapping, non-commuting traffic such as leisure trips may be induced. Drawing on trajectory data from Shijiazhuang, randomly adding traffic during 8-10 a.m. to the traffic at peak hour (7-8 a.m.) after home swapping, we found that the reduction in carbon emissions remains substantial.
At last, we introduce a data-driven model to explore the self-organizing dynamics of residential choice based on urban economics theory. By introducing government coordination to the model, we can obtain a similar result in the reduction in average commuting distance as in GHS and SDGHS, which suggests a feasible approach to achieve a similar residential allocation as in home-swapping.

\section{Results}
Our big data of empirical human trajectory are based on the 4G communication records between base stations and mobile phones provided by one of the three major service providers in Shijiazhuang, the capital of Hebei province in China, from May $22^{\rm nd}$ to $28^{\rm th}$, 2017, and from June $3^{\rm rd}$ to $4^{\rm th}$. The original data include records from around 3 million users out of a total population of 10 million in the city and span 9 days, including 2 weekends and 5 weekdays. To identify meaningful locations in the user trajectories, we divide the whole period covered by the dataset into 15-minute intervals (see Materials and Methods). If an individual stayed in the same location for more than 10 minutes in each interval, the location was recorded as a stop for that user. In addition, when we compute the similarity between the trajectories of individuals later, the division into time intervals facilitates the comparison of defined stops of individuals in each time interval. Other than this high-resolution mobility dataset in Shijiazhuang, we also studied the origin-destination dataset at peak hour (i.e. 7-8 a.m.) of 28 major cities in China to verify the results observed in Shijiazhuang (see Section S1 of the Supplementary Information (SI) for details).

To study the benefit brought by home swapping, one has to obtain two pieces of essential information for each individual, namely (1) his/her location of residence and workplace, and (2) the members in his/her household who also commute. First, we used an individual's trajectory data from the 5 weekdays to infer his/her residence and workplace locations (Fig. 1A). In the subsequent analyses, for each individual, we infer his/her (1) residence location, based on the location he/she stayed for at least four nights in the 5 weekdays, (2) workplace location, based on the location where he/she stayed for at least 2 hours each day and spends the most time during the daytime in the 5 weekdays. Given that traffic congestion in urban central areas is particularly severe, we only studied the main urban area in Shijiazhuang, which contains nearly 420,000 anonymous commuter trajectory records (see Materials and Methods). Although individuals' residence and workplace locations are only inferred, such inference based on cell phone locations should be accurate; this leads to a much larger dataset compared to the data analyzed in conventional studies which are mainly based on surveys.

Next, we infer the household membership between individuals. Since (1) members in the same household usually have different workplace locations but they may enjoy the weekend outing together, and (2) previous studies demonstrated that the intimacy between individuals is related to the similarity between their mobility trajectories~\cite{wang2011human}, we use the similarity between the trajectories of two individuals with the same residence location from the 4 days in the 2 weekends to evaluate their household membership. Specifically, we denote the similarity between the mobility trajectory of individual $i$ and $j$ in a single day by $S_{ij}$ (see Materials and Methods). The closer the value of $S_{ij}$ to $1$, the more similar their trajectories, and closer the relationship between $i$ and $j$. Nevertheless, even in weekends, family members may only spend part of the day together rather than the entire day. Therefore, we determine if $i$ and $j$ are household members by using the highest $S_{ij}$ among the four values found in each of the 4 days of the 2 weekends. If $S_{ij} \ge 0.5$, individual $i$ and $j$ are considered members in the same household. A schematic example of computing the similarity between the trajectories of two individuals is shown in Fig. 1A.

To make our inference more realistic, we ensure the inferred distribution of household size agrees with that found by the census in Shijiazhuang, by grouping individuals $i$ and $j$ with no household members but with higher $S_{ij}$ to form households, more details can be found in Sec. S2 of the SI. The commuting routes between their home and the workplaces of three members of an inferred household are shown in Fig. 1B.

To start analyzing the potential benefit brought by home swapping, we first show the spatial distribution of the location of residence and workplace of commuters in the city in Fig. 2A and Fig. 2B respectively. We divide the urban area of Shijiazhuang into a grid of squares with a dimension of $500 m\times 500 m$; the more the residence or workplace locations in a grid, the higher the intensity of the color on the heat maps in Figs. 2A and 2B. As we can see, areas close to the city center (e.g. Yuhua ,Chang'an ,Xinhua and Qiaoxi Districts) have a high concentration of residence and workplace locations, and the opposite is observed in the surrounding suburban areas. Residence locations are more dispersed and numerous, while workplaces are more concentrated near the city center (see Fig. 2C). The KL divergence (see Materials and Methods) between the two spatial distributions is 0.096, indicating that they are very similar.

Despite the similarity between the overall spatial distributions of residence and workplace locations, it does not necessarily imply that commuters live near their workplaces. As shown in Fig. 2D, the distribution of their daily commuting distance is roughly exponential, such that many commuters commute a long distance daily, suggesting a significant room for exploitation. To examine whether commuters used motor vehicles for commuting, we found that the travel distance of Shijiazhuang residents on-demand vehicles peaked at 2.5 km (see in Fig. S1). We therefore assume that 2.5 km is the shortest distance for commuters to use motor vehicles for commuting. Based on the empirical home-workplace distance, almost $64\%$ of the Shijiazhuang citizens commute a distance larger than 2.5 km and thus they are assumed to use motor vehicles for commuting. In other words, commuters switch from motorized travel to non-motorized travel encompassing walking, cycling, and e-bikes when the commuting distance is less than 2.5 km. This switching threshold of 2.5 km is in line with classification of non-motorized and motorized travel in the 2020 National Commuting Report of China~\cite{2020tongqinbaogao}. In addition, we also shown in Sec. S7 of the SI that different thresholds do not qualitatively affect the obtained results.

In addition, we see that the inferred departure time of commuters from their home peaks within a short period during the morning rush hours in Fig. S2, and together with the long commuting distance, are the dominant factor for urban traffic congestion according to ref.~\cite{wheaton2004commuting,yildirimoglu2021staggered}.

One major reason underlying the long commuting distance for some commuters, is that they live in a location where their other household members have a short commuting distance. We further analyze how households with multiple members choose their home location. In Fig. 2E, we showed the distribution of the difference $\ddiff$ between the largest and the smallest commuting distance among members in the same household. As we can see, the number of households generally decreases with $\ddiff$, with more households found at the small $\ddiff$. The highest peak at the smallest $\ddiff$ implies that many households set their residence location at the centroid among the workplace locations of all the household members, such that $\ddiff$ is minimal.

Interestingly, we note a second peak at an intermediate $\ddiff$, implying that many other households reside at a location closer to one of the household members. As shown in Fig. 2F, dual-commuter households exhibit a clear bifurcation, i.e. some households choose to live in between workplaces such that $(d_{\rm max}, d_{\rm min})$'s lie near the diagonal, while others opt for locations close to one member but farther away from the other with $(d_{\rm max}, d_{\rm min})$'s lying near the horizontal axis. With the increase in the number of household members, it becomes less feasible for families to live close to everyone's workplaces. In this case, we see that $(d_{\rm max}, d_{\rm min})$'s are mainly found near the horizontal axis, implying that they are more inclined to reside in locations close to the workplaces of a few members, sacrificing at least one member's commuting time, consistent with the conclusion in ref.~\cite{surprenant2013commuting}.

\textbf{Home Swapping.} Given the household membership and members' workplaces remain unchanged, we examine the effectiveness of an innovative approach to swap homes among households to reduce the total commuting distance in the whole city. To anticipate the potential benefit brought by such an approach, as shown in Fig. 2A-C, the spatial distributions of home and workplace locations are very similar in the urban area, but most commuters' home and workplace are not in the same location. This means that their home location is someone else's workplace, and their workplace location is someone else's home. As we show in Fig. 2F, it is impossible to find a home location close to every household member's workplace, nevertheless, we still anticipate that after strategic home swapping, the sum of the commuting distance for the swapped households can be reduced, leading to an overall reduction in the average commuting distance at the city level.

\textbf{Random Home Swapping (RHS).} As a benchmark, we randomly swap homes among households and examine its impact on the commuting distance of the city. The results in Fig. 3A shows that the average commuting distance after random home swapping is $2.83$ times higher than that computed based on the real data. It is worth noting that the commuting distance for each commuter in the real data was obtained through crawling the actual commuting path on Baidu Map (see Materials and Methods). This indicates that real households do exhibit an extent of self-organization in choosing home locations, taking into account the commuting distance to each household member's workplaces. Other factors such as transportation, income, amenities around the residence, etc. may also be considered. However, we expect that there is still a large room for improvement at the holistic city level as the empirical data only represent self-organization at the household level.

\textbf{Greedy Home Swapping (GHS).} To examine the benefit brought by strategic home swapping, we propose a greedy strategy to swap homes only if the total commuting distance of the two swapped households decreases. Due to the restriction in living space, we only allow home-swapping between 1-commuter and 2-commuter households, as well as swapping between 3-commuter and 4-commuter households. We terminate the swapping attempts if no households can be swapped to reduce commuting distance for a consecutive $1\times 10^6$ attempts. As the goal of home-swapping is effectively to match each household to a residence location given the workplace locations of all citizens, the Hungarian algorithm is able to identify the optimal solution for this matching problem and hence the optimal allocation of residence locations for all households. Nevertheless, its computation complexity of $O(N^3)$ and the large number of $2.4\times 10^5$ households in our real data would make the matching intractable. Therefore, we employed the above swapping procedure to reduce the commuting distance of the swapped households and the city as whole.

In this case, after home swapping, the decrease in commuting distance fundamentally reduces traffic congestion, through the combined effect of a shorter traveling distance, more dispersed departure times, and a reduced number of motor vehicles used since some commuters may walk or cycle instead. An example of the impact of Greedy Home Swapping (GHS) on individual households is illustrated in Fig. 1C-D. In Fig. 3A, we further show the average commuting distance over all commuters in the city based on the real data, after random and greedy home swapping, which we denote by $\dreal$, $\drand$ and $\dgreedy$ respectively. As we discussed before, $\drand = 2.83\dreal$, and after GHS, $\dgreedy = 0.5\dreal$. To better compare the benefits brought by different home swapping approaches later, we denote $\dreduce$ as the percentage decrease in the average commuting distance over all commuters, given by
\begin{equation}
\label{eq:dreduce}
\dreduce = \frac{\bar{d}_{\rm real} - \bar{d}}{\bar{d}_{\rm real}},
\end{equation}
where $\bar{d}_{\rm real}$ and the $\bar{d}$ are the original and the new average commuting distance over all urban commuters before and after home swapping, respectively. After GHS, $\dreduce = 50.36\%$, which is a highly significant reduction.

In Fig. 3B, we further show the scatter plots of the average commuting distances in a household ($\bar{d}_{\rm household}$) before and after GHS. We find that the average commuting distance for most households are shortened. Based on our empirical data, the mean of $\bar{d}_{\rm household}$ is 8.3 km. After GHS, only $6.1\%$ of households have their average commuting distance increased, and just $0.34\%$ of all households have their average commuting distance increased by more than 5 km.
It is because the households which swap homes with these families have a larger reduction in commuting distances compared to their increases. This is also an example of individual optimum versus social optimum, such that some individual households have to sacrifice to choose a sub-optimal home location to benefit others. Finally, there are also a few households whose commuting distances have not changed, either because they did not swap homes or the commuting distance between their new home and workplaces remain unchanged.

After households' residences are re-allocated, there are changes in their routes, potentially their modes of transportation, and departure time (see Materials and Methods). For instance, commuters with a commuting distance of less than 2.5 km may switch to a non-motorized travel such as bicycles. In this case, the congestion on different road segments at any specific time of the day may change.

To quantify the change in congestion, we define a city-level congestion index, called the average congestion coefficient $I_{t}$ (see Materials and Methods). We then analyze this index at different time of the day based on the real dataset, and after random and greedy home swapping. In Fig. 3C, for all periods of time, we found that the average congestion coefficient after RHS is highest, while that after GHS is lowest, with the empirical one in between. Further analysis was conducted on the traffic condition during the morning peak hour of 7-8 a.m., when most commuters leave home for work. In the case of RHS, most road segments are severely congested and vehicles move slowly. After GHS, only a few road segments remain congested as shown in Figures 3D-F; based on these results, one can anticipate that GHS alleviate congestion throughout the rush hours (see Fig. S4).

\textbf{Greedy Home Swapping with Socio-demographic Consideration (SDGHS).} While Greedy Home Swapping is based solely on reducing commuting distance which significantly alleviates traffic congestion, previous literature has confirmed that socio-demographic factors play a vital role in people's choice of residence~\cite{tillema2010influence,guo2020modeling,rouwendal2001preferences,ho2016workplace,hu2017different,guan2020multiplicity,liu2022quantifying,wang2024does,al2024pathway,cao2017examining,blumenberg2023housing}.

Households, although not fond of spending time in transit, have preference for specific housing characteristics that are strong enough for them to accept a longer commuting time. To identify the difference of each family in capturing economic and locational preferences, we use (1) the distance to the city center, (2) housing price, and (3) amenity accessibility, to serve as proxies to characterize household-level heterogeneity. These metrics were selected because they reflect household-specific attributes rather than purely spatial characteristics, enabling home swapping to take into account the major consideration of individual household while keeping the self-organizing principles of the original GHS framework.

Specifically, on top of reducing commuting distance, two households swap their homes only if
\begin{enumerate}
\item
\textbf{the difference in the distance between their home and the city center is less than $\tc$ km (c-GHS)} - c-GHS portrays the \textbf{\emph{centrality consideration}}  of residential choice. It is because the distance to the city center is an important factor when families choose their homes. According to ref. \cite{xu2023urban} and the data, we define the most visited location in Shijiazhuang to be its city center,  i.e. the largest central business district , i.e. the Beiguo Shopping Mall in Shijiazhuang.
\item
\textbf{the difference between the housing prices of their home is less than $\tp (\%)$ (p-GHS)} - p-GHS incorporates housing price to model the \textbf{\emph{financial consideration}} of households. Since housing price reflects the economic capacity of the household and affect their choice of residence, for instance, they generally choose residence within their financial capacity. Here, we use the city-wide housing transaction data of second-hand properties to describe the average housing price in each region (see Materials and Methods).

\item
\textbf{the difference between the availability of amenities around the two residences is less than $\ta$ (a-GHS)} - a-GHS captures the \textbf{\emph{neighborhood consideration}} via the Points of Interest (POI) density. Since households consider the surrounding living environment when they choose their homes~\cite{abbiasov202415}, here we use the information of POI around the corresponding residence to represent its living environment. For locations without the information of their surrounding POI (e.g. mostly villages), we do not swap their homes with others (see Materials and Methods).
\end{enumerate}

Figures 4A-C show the percentage decrease in the average commuting distance, i.e. $\dreduce$, after Greedy Home Swapping by consider each of above three socio-demographic factors, denoted as c-GHS, p-GHS and a-GHS respectively. We find that as the households' tolerance on the three factors increases, $\dreduce$ also increases, which is reasonable as the socio-demographic consideration limit the flexibility in home swapping.

In Fig. 4D, we further examine $\dreduce$ under the joint consideration of all the three socio-demographic factors with different level of tolerance, which we denote by Socio-demographic Greedy Home Swapping (SDGHS). As we can see, $\dreduce$ reaches $25\%$ under high tolerance, i.e. $\tc = 1 km$, $\tp = 25\%$ and $\ta = 0.5$, but achieves only $3\%$ under low tolerance, i.e. $\tc = 0.1 km$, $\tp = 5\%$ and $\ta = 0.1$. In Fig. 4E, we compare the average congestion coefficient $I_{t}$ computed based on the inferred residence and workplace location from the real dataset and the path of each commuter recommended in Baidu Map during morning and evening rush hours, with those after home swapping under the forementioned socio-demographic consideration with different tolerances. We found that the status of $I_{t}$ under strict consideration (low tolerance) of the factors is closer to that in the real data, while the congestion under loose consideration (high tolerance) has improved as shown in Fig. 4E. $I_{t}$ under medium consideration (medium tolerance), $\tc = 0.5 km$, $\tp = 15\%$ and $\ta = 0.3$, in rush hours is shown in Fig. S5, and over $80\%$ of households benefit under medium tolerance conditions (Fig. S19). For the sake of simplicity in discussion about SDGHS, we used SDGHS with medium tolerance for simulation in our subsequent analysis. In addition to the average congestion coefficient, we analyze the degree of congestion on each road after GHS taking different factors into account, to provide a clearer understanding on how traffic congestion is relieved. As shown in Fig. S6, home swapping can effectively reduce the number of congested and slow-moving road segments during the rush hour. 

In Fig. 4F, it is shown that there is a larger reduction of road congestion between the empirical congestion and that after home swapping under the consideration of housing price and amenity accessibility. However, after GHS taking into account the distance to the city center, some roads remain similarly congested, implying that the distance to the city center is the main consideration of households which hinder the reduction of city-wide commuting distance by GHS. Further comparison among the reductions in average commuting distance based on any 2 out of the 3 socio-demographic factors is shown in Fig. S7, and the results also suggest $\tc$ emerges as the major factor influencing GHS to improving urban congestion.

Furthermore, we examine one important assumption of GHS, such that all households are willing to swap their homes as long as the total commuting distance of the two swapping households decreases, even at the expense of an increase in the commuting distance of one of them. Here, we assume that the households with an increase of commuting distance have an $w\%$ willingness to swap homes but $1-w\%$ to reject. As shown in Fig. S8A of the SI, there is no significant impact on $\dreduce$ as willingness $w\%$ changes.

In the above analysis, we focused on optimizing the commuting distance for households as a whole, without taking into account the individual preference within households. Here, if his/her commuting distance increases, we assume the commuter with an $s\%$ probability refuses to swap homes, which can be considered as individual selfishness in home swapping (see details in Section 3 in SI). Simulation results (Fig. S8B) suggest that the individual selfishness has little impact on $\dreduce$.
However, in reality, residential decisions involve multiple factors beyond the distance to workplace. To take into account any potential requirements of the households, we randomly select a $p$ fraction of households who refuse to relocate regardless of the reduction in commuting distance, and swap homes only for the remaining $1-p$ of households. Simulation results of SDGHS show that even if $10\%$ to $90\%$ of households refuse to swap homes, a reduction of $12.7\%$ to $2.3\%$ of commuting distance is still achieved (see Fig. S9A).

To accommodate heterogeneous individual needs, we incorporate two principal constraints: access to educational facilities for children and the preservation of the shortest commuting distance within each household.
First, considering the population of school-age children in Shijiazhuang, we randomly designate $18\%$ of households to refuse to swap homes, and in this case $\dreduce$ reaches $10.7\%$. To better align with real-world constraints, we analyze the spatial distribution of educational facilities (see Fig. S9B) and define households residing within the average street distance ($272$ m) of any school as ``education-anchored", restricting them from participating in home swapping. SDGHS in this case still achieves a $10.3\%$ reduction in commuting distance (see Fig. S9C).

Second, in another scenario, to account for the household member with the strongest need for a short commuting distance, for instance, caring for young children or the elderly, we refine home-swapping by defining a prioritized member in each household while maintaining system-wide benefits: home swapping are permitted if they (i) reduce the total commuting distance of both households to be swapped and (ii) maintain or reduce the commuting distance of the prioritized member in each household, defined as the one with the shortest commuting distance within the household. In this case, SDGHS achieves a reduction of $8.2\%$ of commuting distance. Under the combined constraints of educational resources and commuting priority, SDGHS still yielded an $8.1\%$ reduction in average commuting distance. All the above results demonstrate the robustness of the benefit of home swapping to real-world heterogeneity.

\textbf{Reduction in Carbon Emissions.} As a by-product, Greedy Home Swapping which reduces the average commuting distance and hence the traffic congestion, also reduces carbon emissions. In general, the longer the commuting distance, the longer the time vehicles stayed on the roads, the more congested were the roads, and the more emissions were produced ~\cite{kissinger2019detailed,xue2022impact,muniz2018urban,tian2024spatial,bohm2022gross,li2023assessing}. To reveal the impact of home swapping on reducing carbon emissions, we examine the total carbon emissions (see Materials and Methods) in various cases as shown in Fig. 5A. As we can see, the total carbon emissions in RHS were much larger than the total empirical carbon emissions, and the total carbon emissions in GHS was the least, such that the reduction in carbon emissions reaches $77.3\%$. We further compare road carbon emissions under RHS, real data, and GHS in rush hours in Fig. S10.

For GHS with socio-demographic consideration, we see in Fig. 5A that carbon emissions are significantly higher, such that the total emissions are higher in c-GHS than those of p-GHS or a-GHS. The total emissions are similar in the case of p-GHS and a-GHS, and the total emissions after SDGHS under medium tolerance are much higher than that with a single constraint. We show the reduction in carbon emissions during the morning and the evening peak hours in each case in Fig. S11, such that similar results are obtained. To more accurately reflect the consideration of households, we also analyze SDGHS with education-anchored households (SDGHS(e)), SDGHS with specific household members with the strongest need for the shortest commuting distance (SDGHS(s)), and SDGHS with the two combined considerations (SDGHS(es)). In these cases, the total emissions can be reduced by $34.4\%$, $27.6\%$ and $27.4\%$, respectively.

As shown in the upper panel of Fig. 5C, after SDGHS with medium tolerance, specifically $\tc = 0.5 km$, $\tp = 15\%$ and $\ta = 0.3$, the carbon emissions of some roads in the city where have decreased significantly compared to the emissions estimated by the original un-swapped data. When we consider only the constraints of $\tp$ and $\ta$, the lower panel of Fig. 5C shows much more roads with reduced carbon emissions. By comparing the insets of the upper and lower panels, we see that the reduction of emissions of the corresponding roads decrease significantly when the distance to the city center is considered in home-swapping in the upper panel, in line with the conclusion in Fig. 4F, implying that the distance to the city center plays a large role on home-swapping in reducing total emissions. 

Next, we analyze the dependence of the reduction in carbon emissions on the reduction in the average commuting distance, i.e. $\dreduce$, after GHS, as shown in Fig. 5B. As $\dreduce$ increases, the emission reduction continuously increases, with GHS reducing $30\%$ of the average commuting distance, leading to a $61\%$ reduction of carbon emissions.
The nonlinear relation between the reduction in carbon emissions and the reduction in commuting distance has several origins. First, the changes in commuters' departure time and route after home swapping change the congestion status of individual road segments. The reduction in congestion decreases the commuting time for a given commuting distance, thereby reduces carbon emissions. Additionally, as we have discussed before, when commuters change their residence locations, their type of travel also changes. When the commuting distance is less than 2.5 km, commuters may switch from motorized travel to non-motorized travel, leading to zero carbon emission. We show the carbon emissions of individuals after GHS in Fig. S12, suggesting a potential trifurcation. Therefore, shortening average commuting distance does not only bring convenience to the residents' daily lives but also hold a strong significance for protecting the urban environment and thus benefiting the health of the urban population.

Based on the above analysis, we found that GHS and SDGHS can effectively reduce the total urban commuting distance, and alleviate the city-wide congestion and carbon emissions. However, considering the heterogeneity of individual, citizens may use public transport to commute a long distance instead of driving. In order to analyze the impact of multi-modal transportation, according to the commuting distance and average commuting speed based on the empirical data, we divide all commuters who use motorized travel into two modes, either by public buses or by private cars. As shown in Fig. S18A, the public transport system is established by the bus route data (see detail in Section S1 of SI). The results of GHS and SDGHS under multi-modal transportation were constructed in Fig. S18B and C, and it is indicated that $\dreduce$ and the reduction of carbon emissions are only slightly affected, meaning the inclusion of public transportation does not compromise the validity of the results of the proposed approach.

While our study primarily focuses on reducing commuting distance through home-swapping, we remark that additional non-commuting traffic may be induced owing to the reduced traffic congestion ~\cite{duranton2011fundamental,metz2021time,loo2022spatio,noland2006flow}. We thus conduct analyses to demonstrate the robust benefit brought by home swapping even with non-commuting traffic induced: we add $0\%$-$25\%$ of trips exceeding $2.5$ km from the dataset during off-peak hours to the improved traffic patterns after GHS (see details in Sec. S8 of SI), such that the reduction in emissions decreases from $78.3\%$ to $71.7\%$ (see Fig. S17). Subsequent to the GHS, we also simulate SDGHS and SDGHS(es) incorporating $15\%$ of induced trips (i.e. roughly $63,000$ motorized trips). As denoted by iSDGHS and iSDGHS(es) in Fig. 5A, the reductions in carbon emissions are $39.8\%$ and $25.5\%$, respectively.
This robust benefit from home swapping comes from the fact that the average distance of motorized commuters is $11.1$ km, higher than the average length of $7.6$ km of induced non-commuting trips. While the above simulation may not fully represent the amount and the pattern of induced trips, the strong periodicity of weekday travel patterns suggests induced trips in weekdays may be limited. These findings confirm the robustness of home swapping benefit and highlight an important direction for future research incorporating activity-based induced travel demand.

\textbf{Exploiting GHS by Polycentric Urban Structure.} As we know, the polycentric city structure affects urban traffic congestion and carbon emissions~\cite{nilforoshan2023human,jun2020effects,zuo2024interactions,rong2022inspection,louail2014mobile,louf2013modeling,pan2024whither,sun2020effects,zhao2024unravelling}. The above analysis also suggests that the distance to the city center is the primary socio-demographic factor affecting the choice of the residence location of households and the reduction of commuting distance by home swapping. We thus study whether a polycentric urban structure could reduce significantly the average commuting distance. In this case, by shifting from a monocentric city model, where most activities are concentrated in a single region, to a polycentric model, with multiple centers of activity across the city. This approach could lead to a more balanced spatial distribution of living-working locations, allowing a more efficient use of transportation infrastructure, reducing travel time and carbon emissions, improving accessibility to amenities and services, and hence enhancing the quality of life for urban residents.

In this analysis, several new activity centers are added to Shijiazhuang in addition to the original city center, each in a relatively bustling area of the city, with a distance of more than 3 km between each pair of centers. We assume that this is a policy-driven scenario, since city centers are planned by the city government at specific locations such that the development of infrastructures will follow, and residents are attracted to these new city centers owing to the expected development.

Specifically, when SDGHS was conducted with medium tolerance, we found that $\dreduce$ increases with the number of city centers. To examine the impact of the distance between the city centers, we study the scenarios when the new centers are the most visited locations outward from the original city center at a minimum distance of 3 km, 6 km, 9 km, and 12 km, respectively, as shown in Fig. 6A. We found that $\dreduce$ increases with the distance between the city centers. For instance, in the case of a five-center city such that the four new centers are established at a distance of 12 km outward from the original center, $\dreduce$ reaches $22.3\%$, suggesting that new urban centers should be established in peripheral areas to achieve a more effective reduction in commuting distance through home swapping. This suggests a strategic approach to urban planning whereas developing multiple activity centers with a distributed pattern and a larger distance between them can significantly improve commuting efficiency by home swapping, hence achieving a more efficient and energy-saving city.

We further select bustling areas in the four peripheral counties of Shijiazhuang as new activity centers, to explore the impact of establishing multiple peripheral centers on the reduction in commuting distance. The results in Fig. 6B show that as the number of activity centers increases, the reduction of commuting distance increases; from a single center to dual centers, even without the corresponding supporting amenities, there is a significant reduction in commuting distance after SDGHS. From dual centers to multiple centers, the marginal increase in the reduction of commuting distance slows down. Other than the reduction in the average commuting distance, we show in Fig. 6C the average congestion coefficient decreases with the number of activity centers, while carbon emissions are reduced after SDGHS. This indicates that home swapping with the establishment of a polycentric city is a promising urban-planning solution in alleviating congestion and reducing carbon emissions.

Finally, we further considered enlarging the amenity accessibility of the newly added activity centers and increasing the housing vacancy based on the original city center (see Fig. S13); conclusions similar to the above were obtained.

\textbf{Universality of the benefit from home swapping.}
To verify the benefit of home swapping in cities other than Shijiazhuang, we studied an additional origin-destination dataset of commuters from another Chinese telecommunication operator, covering the morning traffic peak hour between 7-8 a.m. in 28 major Chinese cities (see Section S1 of SI for details). 
The absence of detailed individual trajectories in this data prevented the direct inference of household membership. Consequently, a simplified household matching mechanism was employed to construct households (see Section S11 of SI). First, we analyzed and found that in each of the 28 cities, the spatial distributions of commuters' residence and workplace locations are similar, as observed in the case of Shijiazhuang, with an average KL divergence of $0.087$ (Fig. S20). This already indicates that home swapping is potentially beneficial in other major cities. As shown in Fig. 7A, we found that the average commuting distances decreases in these cities after GHS.
Additionally, there is a slight positive correlation between $\dreduce$ and the city population size. Notably, Chongqing's KL divergence is $0.067$, and $\dreduce$ after GHS is as high as $88\%$. The abnormally high $\dreduce$ might be caused by the mountainous geography of Chongqing which limit navigation. Likewise, this mountainous geography also makes the use of POI and housing price to describe regional (grid cells) characteristics invalid. To ensure the validity of the experiment, Chongqing is excluded from subsequent analysis.

By incorporating datasets from multiple sources, including amenities (POI data), housing prices, and commuting routes across the 28 cities, we further analyze the effectiveness of GHS with a single socio-demographic factor. In light of the characteristics of the datasets employed, adaptations are essential, involving the calibrated adjustment of key threshold parameters, i.e. the distance to city center with $\tc=3km$, housing price with $\tp=30\%$, and residential amenity accessibility with $\ta=0.3$, to better align the model with empirical conditions. As shown in Fig. 7B, both p-GHS and a-GHS show similar $\dreduce$ and outperform c-GHS. The values of $\dreduce$ of each city, after c-GHS, a-GHS, and p-GHS, are shown in Fig. S21A.

By considering the joint effect of the three socio-demographic factors, Fig. 7C shows the values of $\dreduce$ after SDGHS with $\tc=3km$, $\tp=30\%$ and $\ta=0.3$ for China's top-10 congested cities~\cite{conRep2023}. Although Chongqing ranks second among congested cities, due to Chongqing's special mountainous geography as we mentioned above, Shijiazhuang replaces Chongqing in the following analysis. The results indicate that $\dreduce$ is typically smaller in cities with less severe urban congestion before home swapping, suggesting the significance of home swapping for megacities experiencing severe congestion. Furthermore, we examined the impact of polycentric urban structure on home swapping. Fig. 7E shows that polycentric cities achieved a greater reduction in average commuting distance after SDGHS than their monocentric counterparts (see details in Fig. S21B). To verify the positive correlation between $\dreduce$ and reduction in carbon emissions, we estimated the reduction in carbon emissions after GHS for the 10 cities (see Section S11 of the SI for details). As shown in Fig. 7D, a positive correlation exists between $\dreduce$ and carbon emissions reduction, consistent with the results from the study in Shijiazhuang using high-resolution trajectories (Fig. 5B). Collectively, this validation on multiple cities suggests that our findings do not depend on specific urban context but reflect fundamental principles of reducing commuting distance by home swapping.
It is worth noting that the values of $\dreduce$ obtained by the origin-destination data after GHS are similar to the result obtained by the high-resolution trajectories in Shijiazhuang. However, the $\dreduce$ is significantly higher than that from Shijiazhuang when socio-demographic factors are considered. 
This discrepancy primarily originates from the coarse spatial granularity and the limited temporal resolution of the data, and an overly simplified household matching mechanism (see Section S11 of SI).

\textbf{Data-driven model of residential choice.}
The above analysis shows that home swapping can effectively reduce the city-wide congestion and emissions. To explore the self-organizing mechanism of residential choice, and to analyze the feasibility of government coordination, we developed a data-driven model to describe residential choice (Fig. 8A) based on urban economics theory~\cite{fujita1989urban} and social physics~\cite{yan2014universal,barthelemy2019statistical}.
The model assumes that the attractiveness of location $i$ to household $j$ is dictated by three factors: (1) the housing affordability $E_{i}^{j}$, i.e. the alignment between the income level of household $j$ and the average housing price in the neighborhood of location $i$, (2) the locational utility $U_{i}$ of location $i$, such as amenities and infrastructure scaled by population density, and (3) the geometric mean commuting distance for household members $D_{i}^{j}$. Thus, the attractiveness $A_{i}^{j}$ of location $i$ for household $j$ can be expressed as
\begin{equation}
	A_{i}^{j} \propto \frac{E_{i}^{j}\times U_{i} }{D_{i}^{j}},
\end{equation}
where the expressions of $E_{i}^{j}$, $U_{i}$ and $D_{i}^{j}$ are shown in Materials and Methods. Based on this model and the empirical data of home-workplace location of urban households, we constructed a data-driven residential choice model for individual households. Simulation results show that if households choose their residential location according to this model, the total commuting distance at the city level is $1.1$ times that of the empirical data. Meanwhile, compared with the empirical morning peak average congestion coefficient ($0.75$), the model simulated average congestion coefficient is $0.78$.

To verify the model's validity from the perspective of individual households, we compare the distribution of households' average commuting distances obtained from the model with that of the empirical data. As shown in Fig. 8B, the blue dots represent model simulation results while the red dots show empirical data, demonstrating that the model's statistical distribution aligns well with actual data. Furthermore, we compare the average commuting distance of individual household between model simulations ($\bar{d}_{\rm household}^{Model}$) and the empirical data ($\bar{d}_{\rm household}^{Real}$), as shown in the inset of Fig. 8B. The black dashed line indicates perfect equality between a household's actual average commuting distance and model prediction, while the black solid lines represent cases where the difference between the empirical data and model simulation is less than 10 km. Statistics show that $78.2\%$ of individual households fall between these two black solid lines, indicating the model performs well at the individual household level. Therefore, this data-driven model outlines the principle and theory for analyzing residential location choice in the target city, which performs well at describing the self-organizing mechanism of residential choice.

To examine the compatibility of city-wide home-swapping within the urban economics framework, we extended the original model by incorporating government-coordinated interventions. Specifically, we introduced an economic subsidy policy where the government provides incentives for households to relocate to their guided optimal location of residence, which we call G-residence. The G-residence is defined as the residential location determined for each household through the GHS algorithm, exhibiting significant heterogeneity across different residential locations. Consequently, the relocation subsidies also vary substantially among households, as formalized in the following expression of policy incentive $\alpha_{i}^{j}$ for household $j$ to relocate to area $i$
\begin{equation}
	A_{i}^{j} \propto \alpha_{i}^{j} \times \frac{E_{i}^{j}\times U_{i} }{D_{i}^{j}},
\end{equation}
where the expressions of $\alpha_{i}^{j}$ is shown in the Materials and Methods, location $g$ is the government-recommended G-residence for household $j$ (as shown in Fig. 8A), and $\theta$ denotes the economic subsidy rate (ranging from $0\%$ to $100\%$) that household $j$ would receive when residing at location $g$. A higher $\theta$ value indicates stronger government intervention and greater government fiscal expenditure. To allow feasibility in this incentive provision scheme, we incorporate a distance-decay mechanism for the subsidy. If household $j$ resides in location $i$, they can also have a relative discount, which decays with the distance from location $i$ to G-residence (see details in Materials and Methods).

The reduction of average commuting distance ($\dreduce$) under different subsidy levels is shown in Fig. 8C. As the subsidy intensity $\theta$ increases, the city-wide commuting distance is further reduced. The results clearly demonstrate that in our urban economics-based data-driven residential choice model, with a policy-induced $\theta=10\%$, the total city-wide commuting distance achieves nearly $25\%$ reduction, while the morning peak average congestion coefficient decreases from $0.78$ to $0.55$, corresponding to a $55.46\%$ carbon emission reduction in morning peak hours.

In fact, many households might choose to reside in other location, far from the G-residence, even though the government provides a policy incentive. As shown in the inset of Fig. 8C, where the horizontal axis represents the policy input $\theta$ and the vertical axis shows the actual subsidy expenditure $\theta^{*}$ (i.e., the real financial input under the corresponding $\theta$ values), we observe a nonlinear relationship between $\theta^{*}$ and $\theta$. At $\theta=10\%$, the actual policy expenditure $\theta^{*}$ is only $6\%$. Fig. 8D visualize the city-wide road congestion condition during morning peak hours with $\theta=0\%$ and $\theta=10\%$, along with their respective average congestion coefficients. Comparing to Fig. 3E and Fig. 3F, the results clearly show that the introduction of subsidy for residential choice model achieves a reduction in commuting distance comparable to our proposed approach of city-wide home swapping without subsidy.

\section{Discussion}

In this paper, by analyzing over 400,000 mobility trajectories in the Chinese city of Shijiazhuang, we found the residential choice of households exhibits a self-organizing dynamics in terms of members' commuting distances. Leveraging the observation of approximate distributions of residences and workplaces, we explored the potential benefit brought by a hypothetical home-swapping process between households to reduce traffic congestion and carbon emissions in urban areas. As proof of concept, the first proposed strategy of greedy home swapping (GHS) does not take socio-demographic factors and the needs of individual household members into account to reveal the upper bound of congestion and emissions reduction, and we found that the city-wide average commuting distance (measured by $\dreduce$) could be potentially reduced by over $50\%$. This substantial reduction underscores the critical relationship between the spatial distributions of citizens' residences, workplace locations, and urban traffic congestion.
To increase practical relevance, our analysis incorporated the socio-demographic factors, including the distance to the city center, housing price, and amenity accessibility in swapping homes, serving as proxies to characterize household-level heterogeneity. Compared with the results of GHS, the $\dreduce$ of socio-demographic constrained home swapping (SDGHS) declines but still accounts for $13.4\%$. Notably, among these factors, the distance to the city center emerged as the most influential factor for urban residents in choosing their residence location and deciding whether to swap homes. These results suggest that the coordinated allocation of residence with the development of a polycentric city is an innovative strategy to substantially alleviate peak-hour traffic congestion and carbon emissions in densely populated metropolitans. These findings were supported by a recent study with a small-scale household survey~\cite{zhao2025housing}, while the present study is the first to substantiate such benefits by guiding self-organizing residential choice using large-scale empirical trajectory data.

Nevertheless, both GHS and SDGHS swap homes by considering the household as a unit, neglecting the needs of individual household members. Thus, we further considered the dependence of home swapping on individual needs, especially the availability of schools for children and specific individual household members with the strongest need for a short commuting distance. By defining households within $272$ meters of schools as ``education-anchored" families and restricting them from participating in home swapping, $\dreduce$ declines to $10.3\%$. On the other hand, prioritizing the need for the household members with the strongest desire for the shortest commute, $\dreduce$ reduces to $8.2\%$ after SDGHS. These results underscore the adaptability of our approach to heterogeneous residential preferences. While commuting distance may not always be the top priority, our framework can accommodate different considerations by households in their decision-making process on home swapping.
Other than the reduction of traffic congestion, the combination of shorter commuting distance and less running time on roads (due to reduced urban congestion) leads to significant decreases in carbon emissions, highlighting another important benefit of home swapping. At the city level, without considering the household heterogeneity and individual needs, the reduction in carbon emissions reaches $77.3\%$ after GHS. Constrained by various factors, such as the socio-demographic factors and the needs of individual household members, carbon emissions can still be reduced by $27.4\%$-$34.4\%$.

While home swapping can greatly reduce the distance of commuting trips, the alleviation of traffic congestion may induce new non-commuting trips. By assuming there are $15\%$ new induced trips, and the three socio-demographic factors and the two individual needs are considered during the simulation of home swapping, the carbon emissions reduction reaches $25.5\%$ at the city level. The relation between the reduction in carbon emissions and the reduction in commuting distance reveals a nonlinear relationship, coming from the dual effect of the reduction of city-level commuting distances and street-level congestion.
Development of the public transportation system is beneficial for reducing congestion and carbon emissions, and hence we also examined multi-modal transportation by integrating the city's entire bus network and real-time schedules into our simulations. The results confirm that our proposed home-swapping approaches retain efficacy across diverse transportation modes. The robustness of our approach stems from the inherent flexibility of home-swapping in adapting to local constraints and household preferences, as well as to multi-modal transportation.

To verify the benefit from home swapping in cities other than Shijiazhuang, we used the commuter origin-destination dataset in 28 major Chinese cities to conduct home-swapping simulation. Similar findings on the benefit of home swapping are observed. The consistency between these large-scale simulations with coarse data and our high-precision single-city analysis suggests that our proposed method of home swapping has captured a fundamental urban management principle: coordinated residential choices can greatly maximize benefits at the individual and city levels. This foundational insight could effectively guide future theoretical research and inform policy development aimed at harmonizing micro-level behaviors with macro-scale urban objectives.

Finally, our residential choice model incorporates factors such as housing affordability, proximity to workplaces, and availability of amenities. This model, validated against empirical data, reproduces the observed distribution of households' average commuting distances, thereby bridging our findings with utility-maximization principles. Further, we introduced policy interventions - such as a $10\%$ rent reduction - to explore how targeted incentives can facilitate or motivate the self-organized home swapping. Simulations reveal that even modest policy measures significantly enhance the redistribution of residential choice, accelerating reductions in city-wide congestion and emissions. Through this analysis, we demonstrate that home swapping with incentives of subsidies possesses substantial theoretical feasibility within urban economics theory, which clarifies how our approach complements - rather than contradicts - urban economics theory, offering a practical pathway to integrate sustainability goals into decentralized residential choice dynamics.

We acknowledge the practical challenges of home-swapping on relocation costs and behavioral inertia. However, by demonstrating that voluntary swaps driven by shared economic and environmental incentives can reshape commuting patterns, our findings provide a proof of concept for market-compatible policies (e.g., congestion pricing rebates, green housing tax credits, etc.) that indirectly promote similar self-organization. Our work emphasizes the need for a holistic consideration of urban design, where residential planning is not only about housing city residents but also about integrating socio-economic factors including, amenity accessibility, financial and environmental considerations.
We also remark that the reallocation of residences is a short-term loss, while the reduction in commuting distance is a long-term gain, and hence the gain must outweigh the loss in the long run. In addition, while city-wide home swapping would be costly, our work identifies guiding principles for incremental, incentive-driven policies that align individual and systemic goals. More importantly, our model simulations reveal that even with the modest interventions - such as a $6\%$ effective fiscal outlay via targeted rent subsidies - one could achieve a $25\%$ total commuting distance reduction and a $55\%$ emission reduction by nudging households toward mutually beneficial swaps. These nonlinear returns suggest that partial, phased adoption of home-swapping (e.g., pilot districts, priority corridors, etc.) could yield disproportionate sustainability gains without requiring abrupt, city-wide reorganization. Thus, this study contributes significantly to the existing body of knowledge by providing empirical evidence on a unique, innovative, yet determined home-swapping approach on how urban structure can be coordinated and reformed, specifically towards a polycentric model to save energy, resources, and time by significantly minimizing the fundamental source of traffic congestion. This is particularly relevant to tackling the challenges of both rapid urbanization and increasing urban population and, in a broad sense, sheds light on sustainable city development.

While this study offers important and interesting insights, it has limitations. First, although the major analysis was based on a dataset of high-resolution trajectories obtained in Shijiazhuang and similar results have been observed by using origin-destination data covering 28 major Chinese cities, examining more cities with different cultures and economic backgrounds would improve its generalizability. 
Second, in cities with distinctive geographic complexity, such as Chongqing, conventional evaluation methods become inadequate, potentially compromising the accuracy of an accurate assessment of the benefit brought by home-swapping.
Furthermore, some other special personal preferences, on top of school districts and caregiving needs which we have already included in our study, may affect the choice of residential locations. Some of these factors are challenging to quantify or to obtain due to privacy and the absence of standardized city-level metrics, limiting us from taking these factors into account in our study.  
Additionally, the regional urbanization trajectories, governance structures, local administrative style, and cultural norms (e.g., homeownership preferences) of the target city could also influence implementation. 
Finally, the analysis acknowledges the potential for a rebound effect, where reduced peak-hour commuting traffic may induce additional non-commuting trips, which could offset the benefits of home-swapping. This underscores the critical role of complementary government policies to manage induced travel demand in ensuring the feasibility and effectiveness of home-swapping.

\section*{Materials and Methods}
\textbf{Dataset of Mobility Tracks.}
Our original data is composed of 9-day 4G communication records between base stations and mobile phones served by one of the three major service providers from May $22^{\rm nd}$ to $28^{\rm th}$, 2017, and from June $3^{\rm rd}$ to $4^{\rm th}$.  The original data contain more than 3 million users out of a total population of 10 million in Shijiazhuang, a major city in northern China. There are over 11500 base stations throughout the city, and the position of an individual is recorded as the location of the nearest base station as long as his/her mobile phones communicate with the base stations in 4G. As some phone applications constantly exchange data with back-end servers, the position of individuals can be recorded up to a high frequency of every second. We then divide each single day into $96$ time windows, each with a duration of 15 minutes, and consider a user stops in a location if he/she stays there continuously or discontinuously for 10 minutes or more within the 15-minute time window. Otherwise, if the user does not stay in the same location for more than 10 minutes within a time window, we identify the data point as ``moving".

\textbf{Identifying Individual Residences and Workplaces in Shijiazhuang.} A user's home location is defined as the place where they spend the longest time between 8 p.m. and 7 a.m., and the total duration of stay each night must exceed 6 hours. If these criteria are not met, the location of the home for that individual on that day is considered unknown. The workplace is defined as the location outside the individual's home, where he/she spends the most time between 7 a.m. and 8 p.m., and the total duration of stay each day must exceed 2 hours. If these requirements are not met, the workplace data for that individual on that day is empty.
Given five weekdays of data, a maximum of five home location assessments can be made for one individual. To mitigate the impact of base station spacing, base stations with home locations less than 250 meters apart are merged, using the location where the longest stay occurred as a substitute. Then, the individual's nighttime data is updated to obtain the position of his/her home each night. For an individual with five days of home location assessments, the location is considered a clear residential characteristic only if at least four days point to the same place. Similarly, there are five records for the workplace. We define the workplace for an individual as the location that appears most frequently among the five assessments. If two locations occur with the same number of times, the total duration of stays over the five days at both locations is compared, with the longer duration determining the individual's workplace.

In our study, considering that urban congestion is primarily concentrated in the central areas of the city, we focus on the main urban districts in Shijiazhuang, including Yuhua, Xinhua, Chang'an, Qiaoxi, Zhengding, Gaocheng, Luquan and Luancheng. Only individuals live or work within these main urban districts are considered, resulting in a trajectory dataset from 422,454 anonymous commuters.

\textbf{Similarity between Individual Trajectories.}
The daily similarity $S_{ij}$ between individual $i$ and $j$ is given as follows,
\begin{equation}
	\label{eq_similarity}
	S_{ij} = \frac{1}{L_{ij}}\sum_{t=1}^{M}\Theta(\theta_c-\theta_{ij}(t))\Theta(d_c-d_{ij}(t))
\end{equation}
where the step function $\Theta(x) = 1$ if $x\ge 0$ or otherwise $\Theta(x) = 0$. The variable $\theta_{ij}(t)$ is the angle between the residence and the position of individual $i$ and $j$ at time $t$; $\theta_c = 30^{o}$, while $d_{ij}$ represents the distance between individuals $i$ and $j$ at time $t$ and $d_c = 272 m$, which is the average street length in Shijiazhuang obtained from OpenStreetMap (see SI section 1). $M=96$ is the total number of time intervals in one day, and $L_{ij}$ represents the number of time intervals when the location information of both individuals $i$ and $j$ are available and they were not at their home. We require $L_{ij}\ge 4$ for $S_{ij}$ to be considered.

\textbf{The KL Divergence between the Spatial Distributions of Residence and Workplace.}
The KL Divergence between spatial distributions of residence and workplace is calculated as follows,
\begin{equation}
	KL(p||q)=\sum_x p_x\log\frac{p_x}{q_x}
\end{equation}
where $p_x$ and $q_x$ denote the density of residence and workplace in grid $x$, respectively. A smaller KL divergence indicates a smaller degree of difference between the two distributions.

\textbf{Inferring Commuting Route and Home-Departure Time.} To obtain the actual commuting distance before and after commuters swap their homes, we combined different residence and workplace locations in pairs and crawled the routes for each Origin-Destination (O-D) pair via the navigation function on Baidu Map. Due to the large volume of data, we grouped different locations to the nearest road center. If the residence and workplace are on the same road, we used the same location of origin. Each data entry includes the residence, workplace, commuting distance, commuting route, and the time spent on each section of the route.

The inferred home-departure time is the average time each commuter first leaves home between 5 a.m. and 11 a.m. on the 5 weekdays covered by the data. Commuters not meeting this criterion are assigned departure time proportionally. As shown in Fig. S2 of the SI, the proportion of commuters departing home between 7 a.m. and 8 a.m. is the highest. The home-departure time after home swapping is inferred as the time the commuter has to start their journey in order to reach their workplace at the same time before home swapping. The distributions of home-departure time after GHS and SDGHS are shown in Fig. S3A and S3B of the SI.

\textbf{Average Congestion Coefficient.} To explore the impact of shortened commuting distances on traffic congestion after home-swapping, we define an average congestion coefficient $I_t$ as follows
\begin{equation}
I_t = \bar{c_{\alpha}^t} = \frac{\sum_{\alpha=1}^{M_t} c_{\alpha}^t}{M_t},
\end{equation}
where $M_t$ is the number of roads in the set of congested and slow-moving segments at time $t$ in empirical situation, and $c_{\alpha}^t$ is defined as the congestion coefficient for each road segment~\cite{de2014determining}, given as follows
\begin{equation}
c_{\alpha}^t=\frac{N_{\alpha}^t}{\tilde N_{\alpha}}  ,
\end{equation}
where $N_{\alpha}^t$ is the number of vehicles traveling in the same direction on road segment $\alpha$ at time interval $t$, such that we calculate the time of commuters starting on each segment based on their travel time and departure time. It is important to note that the time point is calculated hourly collected from four 15-minute intervals, and commuters using non-motorized vehicles have been excluded. If a commuter passed road $\alpha$ during the time interval $t$, the volume of the road $\alpha$ at time interval $t$ is incremented by one. Hence, the real motor vehicle volume on every road in each time interval is accurately counted. It is stipulated that the morning arrival time at workplace and evening departure time from workplace remain unchanged. Thus, after GHS, the morning departure time and evening arrival time from the new residence can be deduced from navigation data, such that the new motor vehicle volume on every road can be estimated in any time interval. $\tilde N_{\alpha}$ is the maximum traffic volume of the road $\alpha$, given by
\begin{equation}
\tilde N_{\alpha} = \frac{1000 \times \tilde v_{\alpha}}{l_{c}+l_{a}^{\alpha}},
\end{equation}
which varies depending on the speed limit $\tilde v_{\alpha}$ of road segment $\alpha$(see navigation data in SI Section 1). $l_c$ is the average length of a vehicle, i.e. 4 meters; $l_{a}^{\alpha}$ is the length of the safety distance between vehicles, which is dependent on the maximum driving speed of road $\alpha$, e.g. $l_{a}^{\alpha} = 100 m$ if speed limit $\tilde v_{\alpha} = 100 km/h$~\cite{mcshane1990traffic}. In consideration of the impact of multi-modal transportation, the city's entire bus network was integrated. The average length of automobiles is  $l_c$, whilst the average length of buses is four times of that of automobiles.

\textbf{Identifying Housing Price.} We identify the price of a housing unit at a residence location based on their price listed in the second-hand market from the website \url{Lianjia.com} for Shijiazhuang,  where information such as the name of the properties, the latitude and longitude of the residential communities, their total price and the unit price, are available (see SI Section S1). We have discretized the city into $0.005^{\circ}$ ($\sim 0.5 km \times 0.5 km$) grid cells and identified the housing price $/sq.m$ averaged over all transacted residences for each individual cell, capturing localized housing costs while accounting for spatial heterogeneity. For grids without housing price data, the price of a housing unit is assumed to be the average of the housing prices of the nearest four grids.

\textbf{Inferring the Dissimilarity in Amenity Accessibility.} We utilize the data of point of interest (POI) crawled from Gaode, a map app used in China, which includes each POI's name, type, location, area, and street information. We then select seven major categories of amenities: (1) dining, (2) shopping, (3) transportation facility, (4) science, education and culture, (5) services, (6) sport and leisure, and (7) medical and security, such that the weighted proportion of the $k$-th type of amenities at location $x$ is given as follows,
\begin{equation}
a_{x}^{(k)} = \frac{n_{x}^{(k)}}{\sum_{k=1}^7 n_{x}^{(k)}}\times\frac{N^{(k)}}{\sum_{k=1}^7 N^{(k)}},
\label{e:poi}
\end{equation}
where $n_{x}^{(k)}$ and $N^{(k)}$ denote the number of the $k$-th type of amenities, at base station $x$ and across all studied locations, respectively. On top of the amenities right at their residential location, residents can utilize amenities nearby. We thus define the amenity accessibility of the $k$-th type of amenities at location $x$ as follows,
\begin{equation}
A_{x}^{(k)}=\sum_{y} a_{x}^{(k)} \Theta(d_{xy}-L)\left[1-\left(\frac{d_{xy}}{L}\right)\right]^{2},
\label{e:En}
\end{equation}
where $\Theta(x)$ is again the step function such that $\Theta(x) = 1$ when $x \leq 0$ or otherwise $\Theta(x) = 0$; $L$ is the average street distance in Shijiazhuang, which is determined from Open Street Map (OSM) data to be $L=272m$. Finally, we calculate the residential similarity between location $x$ and $y$ by using the cosine similarity between $\vec{A}_x$ and $\vec{A}_y$, where $\vec{A}_x$ is the vector composed of elements $A_{x}^{(k)}$. The amenity dissimilarity $D_{xy}$, which describes the difference of the accessibility to various amenities between location $x$ and $y$, is given by
\begin{equation}
D_{xy} = 1- \frac{\vec{A}_x \cdot \vec{A}_y}{|\vec{A}_x| |\vec{A}_y|},
\end{equation}

\textbf{Urban Carbon Dioxide Emissions}
For exploring how GHS affects carbon dioxide emissions from the whole city, we define the carbon dioxide emission $E(t)$ during each time interval $t$ as in ref.~\cite{kissinger2019detailed}
\begin{equation}
E(t) = \sum_{i=1}^{N_{c}} \sum_{\alpha\in Path_{i}(t)}d_{i}^{\alpha} \times \zeta \times c_{\alpha} \times S_{\alpha} \times R_{\alpha},
\end{equation}
where $N_{c}$ is the number of commuters traveling at time interval $t$; $Path_{i}(t)$ is the set of road segments commuter $i$ passes during time interval $t$, and $d_{i}^{\alpha}$ is the length of road segment $\alpha$; $\zeta$ is the carbon dioxide emission factor~\cite{zhou2022identifying}, which is equal to 0.209 $kg CO_{2} e /km$; $c_{\alpha}$ is the number of vehicles passing road $\alpha$ at the same time with commuter $i$; $S_{\alpha}$ and $R_{\alpha}$ represent the slope and the type of road $\alpha$, which are both set as constants for simplicity. In addition to the total carbon emissions, we can also obtain the emissions from each commuter or on each road in each time interval. Additionally, given that the buses use diesel, the carbon emission factor is 0.281 $kg CO_{2} e/km$.

\textbf{Data-driven model of residential choice.} We construct the data-driven model of residential choice based on households' preference as following
\begin{equation}
	A_{i}^{j} \propto \frac{E_{i}^{j}\times U_{i} }{D_{i}^{j}},
\end{equation}
here, $E_{i}^{j} = 1- \frac{\left |B^{j}-R_{i}  \right | }{B^{j}}$. $R_{i}$ represents the rental price of location $i$, which is proportional to the housing price of location $i$; $B^{j}$ represents the observed rental expenditure of household $j$ (economic capacity proxy). Residential choice probability follows an inverse exponential relationship with price disparity $\left |B^{j}-R_{i}  \right |$, reflecting households' preference for residence matching their financial capacity.

In Henderson's urban system model, there exists an inverted-U shaped relationship between regional scale and residents' regional utility~\cite{henderson2014economic}. Based on this theory, we assume that the attractiveness of a residential location is proportional to the utility value of that region. The utility of location $i$ for household $j$ is related to the population size of the region: when the population size of location $i$ is small, all resident households can benefit; but as the population size increases, the per capita accessibility to infrastructure amenities begins to decline, and consequently, the probability of household $j$ in choosing location $i$ as residence decreases. Therefore, we define the impact of location $i$'s locational utility ($U_{i}$) on household $j$'s choice as $\sqrt{(S_{i}-O_{i})(O_{i}+k)}$, such that $S_{i}$ represents the total population capacity of location $i$ as a residential area, $O_{i}$ represents the current population size of location $i$ and $k$ is the initial parameter, which is set to $2$ in our model.

The attractiveness of area $i$ for household $j$ is formulated as follows
\begin{equation}
	A_{i}^{j} \propto \left( 1-\frac{\left| B^{j}-R_{i}  \right|}{B^{j} } \right) \left( \frac{\sqrt{(S_{i}-O_{i})(k+O_{i})} }{\sqrt[m]{ {\textstyle \prod_{p=1}^{m}d_{iW}^{C^{j}_{p}}}}}\right),
\end{equation}
where $d_{iW}^{C^{j}_{p}}$ represents the commuting distance of the $p$-th commuting members in household $j$ if they reside at location $i$, and $m$ denotes the number of commuting individuals in the household. To balance the influence of each family member's commuting distance, we characterize this using the geometric mean of family members' commuting distances, i.e., $D_{i}^{j} = \sqrt[m]{{\textstyle \prod_{p=1}^{m}d_{iW}^{C^{j}_{p}}}}$. To prevent division by zero, the minimum commuting distance for any family member is set to $1 m$ in the simulation.

Considering the restriction in living space, all the households could be divided into two groups: single/double-commuter households, and triple/quadruple-commuter households. Households in the same group can exchange their residences, but not between groups, i.e., 1-commuter households and 2-commuter households can exchange their residences, as well as between 3-commuter and 4-commuter households. As the number of households of each group at location $i$ remains unchanged, the total number of residents may be more than the empirical capacity of location $i$. Therefore, $S_{i}$ can be defined as the numerical upper bound, which is equal to the sum of twice the number of single/double-commuter households and four times the number of triple/quadruple-commuter households.

The G-residence is defined as the residential location determined for each household through the GHS algorithm, exhibiting significant heterogeneity across different households. Consequently, the subsidy also varies substantially among households, as formalized in the following theoretical expression
\begin{equation}
	\alpha _{i}^{j} \propto \left [ 1-\frac{\theta}{(1+\frac{d_{iG} }{L} )^{2} }   \right ] ^{-\delta },
\end{equation}

For household $j$, $\alpha_{i}^{j}$ represents the incentive provision by government policy, where location $g$ is the government-recommended G-residence for household $j$ (as shown in Fig. 8A), and $\theta$ denotes the economic subsidy rate (ranging from $0\%$ to $100\%$) that household $j$ would receive when residing at location $g$. To allow feasibility in this incentive scheme, we incorporate a distance-decay mechanism for the subsidy, where $L$ is the average street distance in the target city and $d_{iG}$ is the distance between location $i$ and the G-residence. The smaller $d_{iG}$, the closer the subsidy amount at location $i$ to $\theta$ for household $j$. The parameter $\delta$ represents the household's sensitivity to subsidy incentives - higher values indicate more responsive behavior from household $j$ to the policy incentives. In our simulations, this parameter is set to $\delta=4$.

\clearpage
\noindent \textbf{Acknowledgments.} \\
C.Z. acknowledges the Natural Science Foundation of Hebei (No. F2020205012), the Youth Top Talent Project of Hebei Education Department (No. BJ2020035), and the Project Supported by Science Foundation of Hebei Normal University (No. L2023K04). The work by C.H.Y. is supported by the Research Grants Council of the Hong Kong Special Administrative Region, China (Projects No. EdUHK GRF 18301217, and No. GRF 18301119), the Dean's Research Fund of the Faculty of Liberal Arts and Social Sciences (Projects No. FLASS/DRF 04418, No. FLASS/ROP 04396, and No. FLASS/DRF 04624), and the Internal Research Grant (Project No. RG67 2018-2019R R4015 and No. RG31 2020-2021R R4152), The Education University of Hong Kong, Hong Kong Special Administrative Region, China.

\noindent \\ \textbf{Author contributions.} \\
Y.-Q.L., C.Z., C.H.Y. and A.Z. designed the research, Y.-Q.L., C.Z. , X.-Y.Y., X.H. performed the experiments, Y.-Q.L., C.Z., X.-Y.Y., C.H.Y. and A.Z. analyzed the data, C.Z., C.H.Y. and A.Z. wrote the paper.\\

\noindent \textbf{Competing financial interests.} The authors declare no competing financial interests.\\

\noindent \textbf{Data and materials availability.}  The raw data cannot be shared by the authors due to the Mobile Privacy Policy of China, but might be available upon reasonable request submitted to the mobile network operator (China Mobile) at http://it.10086.cn/services/$\#$dsjArea.


\bibliographystyle{unsrt}
\bibliography{References}


\clearpage
\section*{Figures}
\begin{figure}[h!]
\centering
\includegraphics[width=15.5 cm]{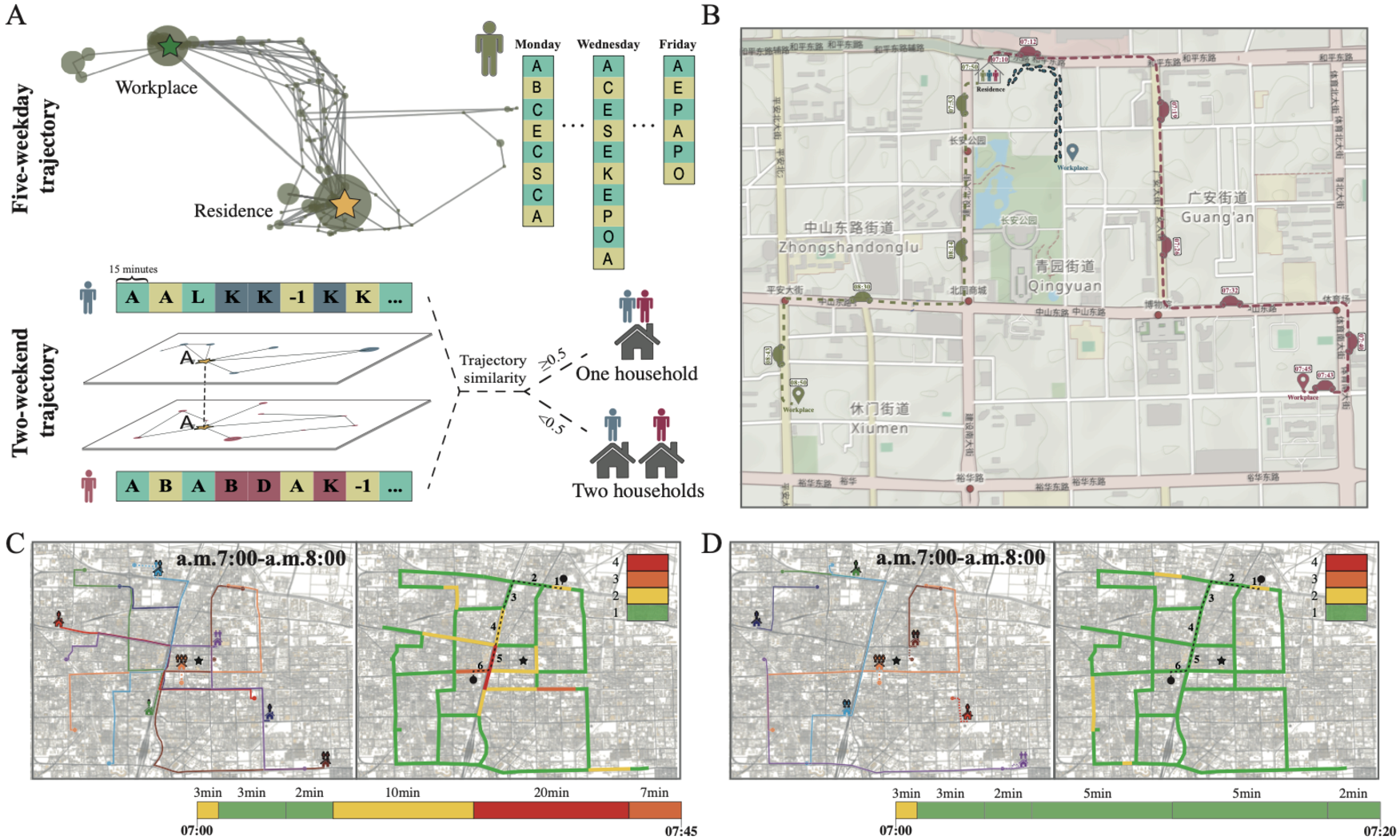}\\
\caption{\textbf{Schematic illustration of urban congestion before and after home swapping in Shijiazhuang.} (A) An individual mobility trajectory in a weekday (upper panel); circles represent visited locations, with their size corresponds to the visit duration. His/her workplace in the daytime and residence by night can be identified, with detailed daily visitation sequence. The mobility trajectory of two individuals in weekends (lower panel), with movement patterns and visit duration, for the inference of household membership based on the similarity between the trajectories. (B) The commuting routes between home and workplaces based on the Baidu map of three members in a household, showing the difference between motorized (red and green dashed-lines) and non-motorized (black dashed-lines) travel based on commuting distance. (C) The morning peak hour commuting routes from 7 households and the corresponding road congestion level. (D) The new commuting routes from the 7 households in (C) after home swapping and the corresponding reduced congestion level. The dashed line represents the same trip during 7-8 a.m., and the color bars show the different congestion situations between (C) and (D).}\label{fig1}
\end{figure}

\begin{figure}[h!]
\centering
\includegraphics[width=16 cm]{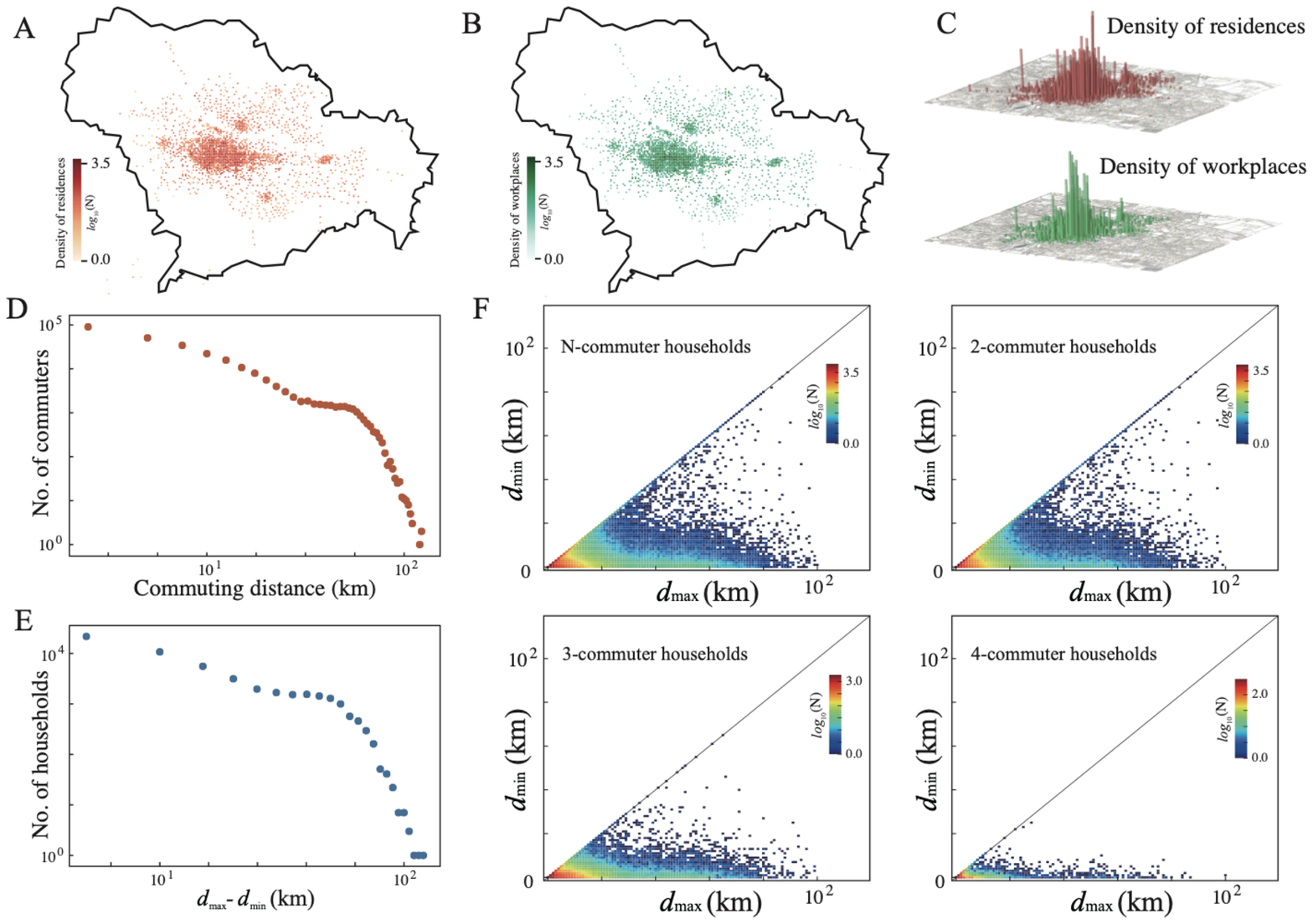}\\
\caption{\textbf{Residential and workplace distribution and the residents commuting behaviour in Shijiazhuang.} (A) The spatial distribution of residences; (B) The spatial distribution of workplaces; (C) A comparison between the spatial distributions of residences (top) and workplaces (bottom); (D) The distribution of commuting distance of Shijiazhuang commuters in log-linear plot. (E) The distribution of $d_{\rm max}-d_{\rm min}$, i.e. the difference between the maximum and minimum commuting distance among members in the same household; (F) The scatter plot of $(d_{\rm max}, d_{\rm min})$, i.e. the maximum and minimum commuting distances among members in the same household, from left to right, for all households, dually, triply, quadruply employed households, respectively.}
\label{fig2}
\end{figure}

\begin{figure}[h!]
\centering
\includegraphics[width=16 cm]{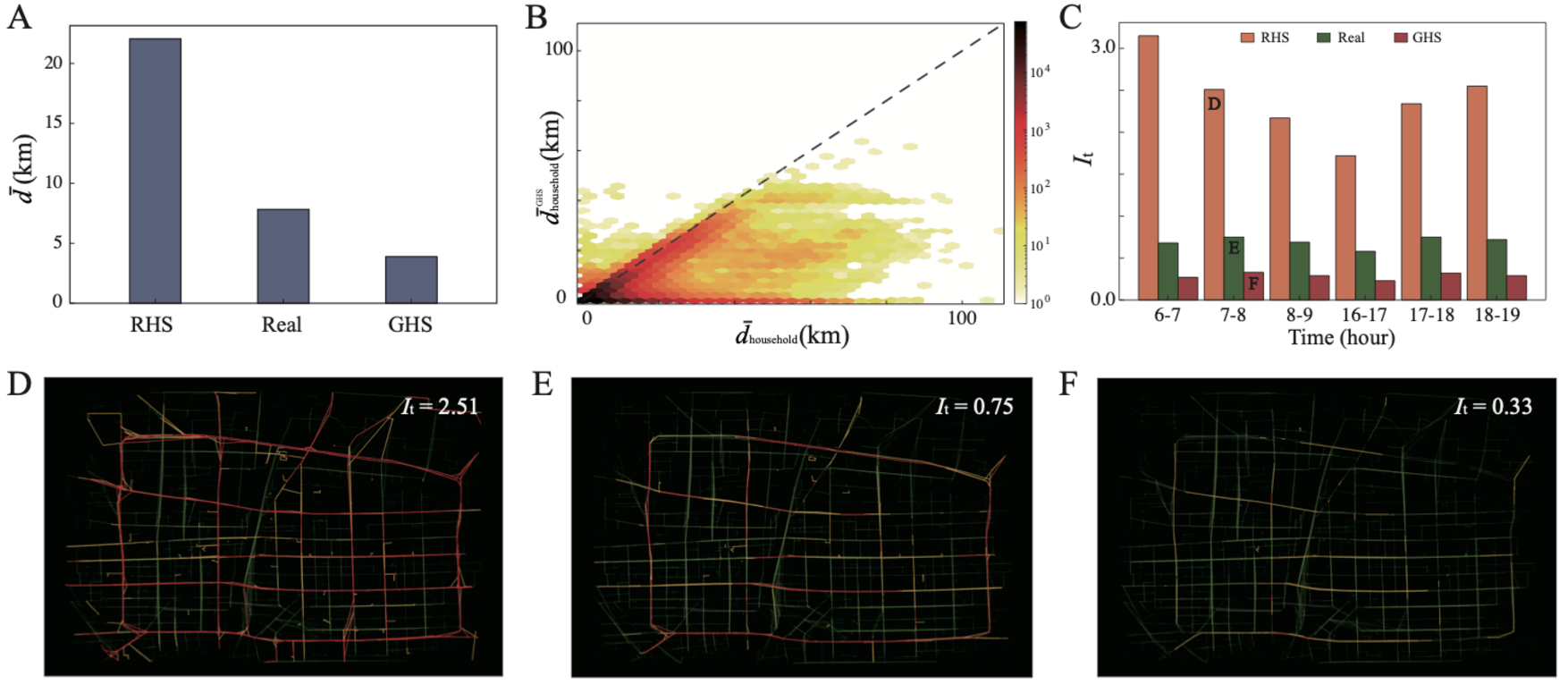}\\
\caption{\textbf{The benefits from greedy home swapping.} (A) The average commuting distances over all urban commuters in Shijiazhuang, after random home swapping (RHS), based on the real data, and after greedy home swapping (GHS); (B) The scatter plot of the original and the new commuting distance averaged over members in a household before and after GHS; (C) The average congestion coefficients during the morning and the evening peak hours based on the real data, after RHS and GHS; (D) - (F) Snapshots of the traffic condition (D) after RHS, (E) based on real data, and (F) after GHS respectively, during 7-8 a.m. in Shijiazhuang.}\label{fig3}
\end{figure}

\begin{figure}[h!]
\centering
\includegraphics[width=16 cm]{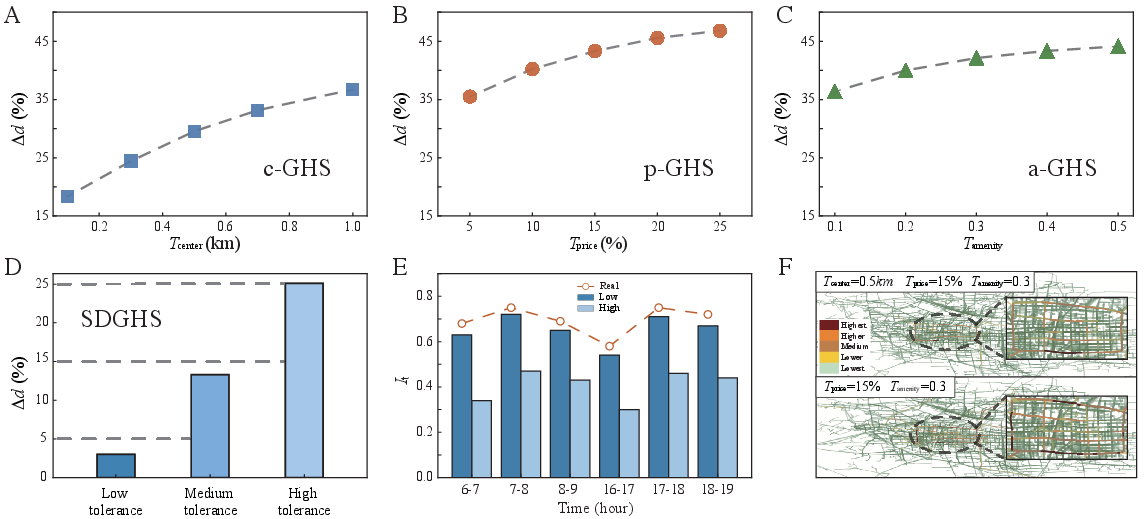}\\
\caption{\textbf{Greedy home swapping with socio-demographic consideration (SDGHS).} (A) - (C) The percentage decrease in average commuting distance, i.e. $\dreduce$, after home swapping with consideration on (A)the distance to the city center (c-GHS); (B) housing price (p-GHS); (C) amenity accessibility (a-GHS); (D) $\dreduce$ after SDGHS with different levels of tolerance; (E) The average congestion coefficients $I_{t}$ at different time based on the real data, and under low tolerance and high tolerance; (F) The difference from the actual road capacity under the joint consideration of the three factors with medium tolerance (upper panel), and housing price and amenity accessibility with medium tolerance (lower panel) during 7-8 a.m. traffic peak.}\label{fig4}
\end{figure}

\begin{figure}[h!]
\centering
\includegraphics[width=16 cm]{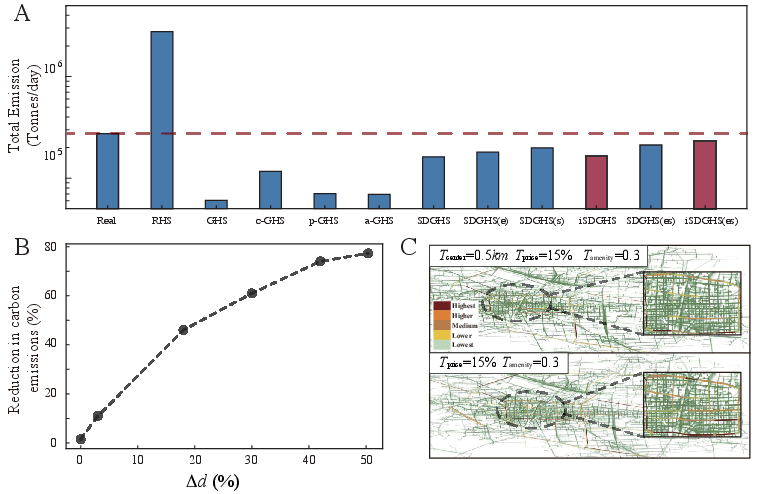}\\
\caption{\textbf{Reduction of carbon emissions by home swapping.} (A) The total emissions in different situations, with iSDGHS and iSDGHS(es) representing the results with induced traffic. (B) The relationship between $\dreduce$ and emission reduction after GHS. (C) The difference from the actual road emission under the joint consideration of the three factors with medium tolerance (upper panel), and housing price and amenity accessibility with medium tolerance (lower panel) during 7-8 a.m. traffic peak.}\label{fig5}
\end{figure}

\begin{figure}[h!]
\centering
\includegraphics[width=16 cm]{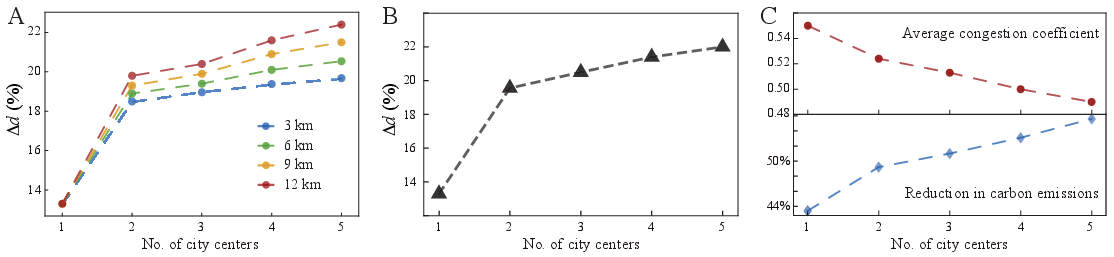}\\
\caption{\textbf{Home swapping with polycentric city structure.} (A) The dependence of $\dreduce$ on the number of newly established city centers in $n$ most visited locations according to the empirical data,  with a restriction the new centers are at least 3 km, 6 km, 9 km, 12 km apart from original city center; (B) The dependence $\dreduce$ on the number of newly established city centers in peripheral counties of Shijiazhuang; (C) The impact of adding new centers, which are established at the four most visited location in peripheral counties, on average congestion coefficient and reduction of carbon emissions (7-8 a.m.). All results were obtained by SDGHS under medium tolerance.}\label{fig6}
\end{figure}

\begin{figure}[h!]
	\centering
	\includegraphics[width=16 cm]{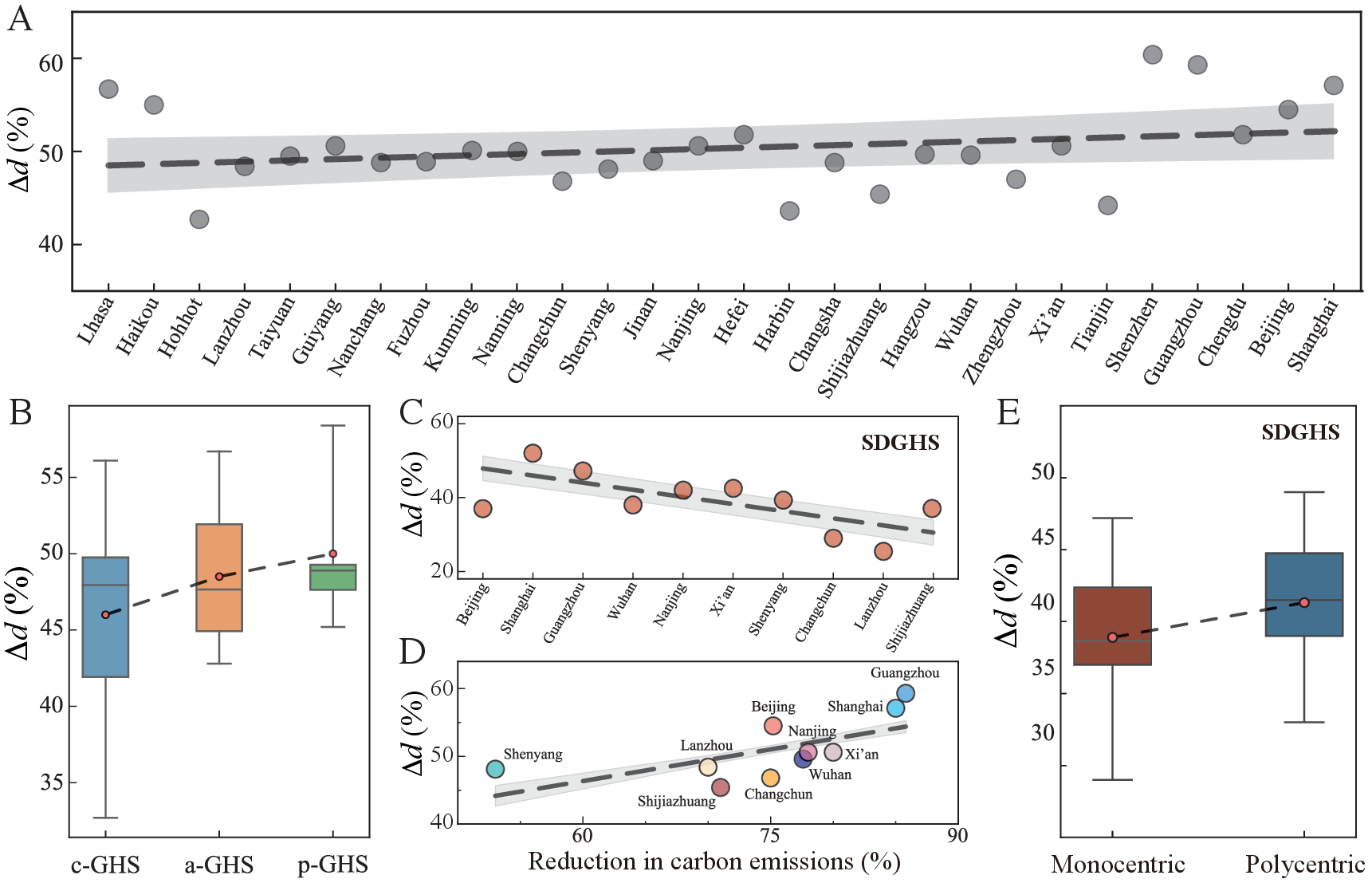}\\
	\caption{\textbf{Universal optimization benefits of home swapping under varying constraints in 28 major cities of China.} (A) the performance of GHS in China's 28 major cities (ordered by population), (B) considering a single socio-demographic factor, the optimization of GHS in China's 28 major cities and (C-E) Chinese top-10 congested cities. (C) the $\dreduce$ after SDGHS in top-10 congested cities. (D) the relationship between $\dreduce$ and the reduction in carbon emissions in the top-10 congested cities after GHS. (E) the comparison of $\dreduce$ between monocentric cities and counterpart polycentric (five-center layouts) cities after SDGHS. All results represent averages from ten independent simulations per city, with shaded areas indicating $95\%$ confidence intervals.}\label{fig7}
\end{figure}

\begin{figure}[h!]
	\centering
	\includegraphics[width=16 cm]{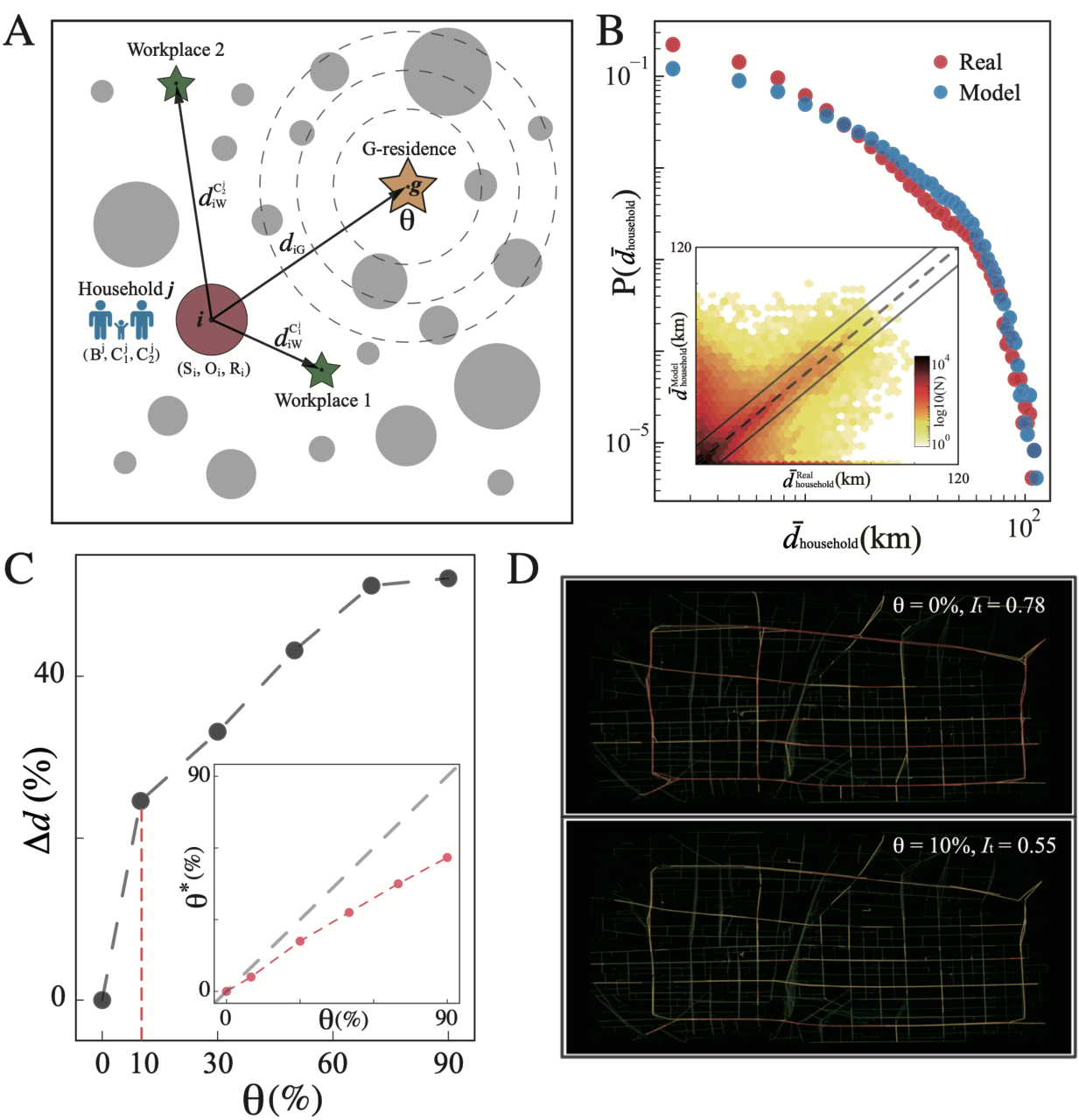}\\
	\caption{\textbf{Data-driven model of residential choice.}
	(A) Schematic of the residential choice model. Household $j$ ($3$ members including $2$ workers at workplace 1 and 2) evaluates residential location i with total capacity $S_{i}$, current population $O_{i}$, and rent $R_{i}$. $d_{iw}^{C_{1}^{j}}$ and $d_{iw}^{C_{2}^{j}}$ represent the commuting distance of the two working members if this household chooses location $i$. G-residence shows the policy-recommended optimal location, and $\theta$ denotes the economic subsidy rate that household j would receive when residing at G-residence, and $d_{iG}$ being the distance between $i$ and G-residence. (B) The comparison of the distributions of the households' average commuting distances between empirical data (red) and model results (blue). The inset depicts the deviation between the simulated and empirical average commuting distances of individual households. (C) The $\dreduce$ under different subsidy levels $\theta$. Inset shows relationship between nominal subsidy $\theta$ and actual expenditure $\theta^\ast$. (D) The city-wide road congestion patterns under $\theta = 0\%$ (upper panel) and $\theta = 10\%$ (lower panel) subsidy scenarios.\label{fig8}}
\end{figure}

\clearpage

\begin{center}
{\large\bfseries Supplementary Information}\\[8pt]
{\large Guiding Self-Organizing Dynamics of Residential Choice in Cities to Reduce Traffic Congestion and Carbon Emissions}\\
\small Yu-Qing Liu, Chen Zhao, Xiao-Yong Yan, Xiaoyue Hou, Chi Ho Yeung, An Zeng
\end{center}
\tableofcontents

\clearpage

\section{Data Description}
The details of the 12 independent datasets used in this study are given as follows:

\begin{itemize}
\item Shijiazhuang Individual Mobility Trajectory Dataset~\cite{trajectory_data2023}: The empirical mobility trajectory big data used in this article is based on 4G communication records between cell towers and mobile phones provided by one of the three major service providers in Shijiazhuang, a major city in northern China, from May 22 to 28 and June 3 to 4, 2017. This dataset covers a total of 9 days (5 weekdays and 4 weekend days). To identify meaningful locations in users' mobility trajectories, the dataset was divided into 15-minute intervals, 96 intervals per day. If an individual stays at the same location for more than 10 minutes in one 15-minute interval, the user is considered to have stayed in that location.

\item Shijiazhuang Commuting Route Navigation Data: To obtain the actual commuting distances before and after commuters swap their homes, first, different residential and workplace locations are paired. Due to the large volume of data, in this study we aggregate different location points to the nearest road center. If the residence and workplace are on the same road, the original location points are used. In this study, we used the aggregated locations to crawl the driving routes for each pair of commuting locations on the Baidu map~\cite{platform_bd}, during the morning traffic peak (6:00 A.M.-10:00 A.M.) and early evening traffic peak (4:00 P.M.-8:00 P.M.) in weekdays, which includes the time taken for an individual to travel through each road segment, the type and the distance of the road segment, and navigation details. The final data include the individual's starting location, the aggregated starting location, commuting distance, commuting time, and commuting route.

\item Shijiazhuang On-demand Vehicle Dataset: The data from DiDi on-demand vehicles in Shijiazhuang for five weekdays in July 2017, including the pick-up and drop-off location of each trip.

\item Shijiazhuang Housing Price Data: The information on second-hand housing prices from Lianjia.com~\cite{lj_sjz} within the Shijiazhuang city area in 2017, including total price, unit price, community name, and community location.

\item Shijiazhuang Point-of-interest (POI) Data~\cite{poi_data}: The information on various types of Points of Interest (POI) within the Shijiazhuang city area collected from Gaode Map in 2017, including store name, type, location, area, and street information.

\item Shijiazhuang Urban Road Network Data: Obtained from OpenStreetMap~\cite{OSM} to facilitate aggregating residence or workplace locations to the nearest road center. The total number of road segments is 28.9 million. The average street length in the city, calculated from crawling the length of streets at all levels, is found to be 272 meters.

\item Shijiazhuang bus route data: All bus routes in Shijiazhuang were crawled from the public transportation website~\cite{Bus}, and the total number of bus routes is 246, and the total number of bus stops is 3661. The data include bus name, bus ID, starting station, terminal station, and passing bus station.

\item Shijiazhuang bus departure schedule data~\cite{Bus}: The data include bus name, starting station, terminal station, first and last departure time.

\item Urban Mobility Dataset of Chinese Major Cities~\cite{MajorCities_data2019}: The aggregated and anonymised mobile phone dataset was provided by a Chinese telecommunications operator. The raw dataset consists of mobile phone signal records collected over a two-month period, representing approximately 90 million residents across 60 cities. The number of anonymised users represents, on average, $18\%$ of the total population and corresponds to the proportion of mobile phone users to the overall population across the 60 cities studied. The operator partitioned all cities into a 1 km $\times$ 1 km grid cell. The aggregated, non-personally identifiable human movement records represent the number of trips between grids per hour in August and November 2019 in 60 Chinese cities. For characterizing the commuting behavior of residents, using data only covering the time span between 7:00 and 7:59, which corresponds to the typical morning peak hour.

\item Chinese Major Cities' Commuting Route Navigation Data: Residence and workplace pairs were first defined by linking trip origins and destinations. Due to the 1 km grid resolution, the precise latitude and longitude for each location were represented by the coordinates of the grid center. Driving routes for each pair were then collected from the Baidu Map API~\cite{platform_bd}. The collected data include travel time for each road segment, road type, segment distance, and detailed navigation information. The final dataset is structured identically to that of Shijiazhuang.

\item Chinese Major Cities' Housing Price Data~\cite{fang_data}: The information of second-hand housing prices from national residential estate historical transactions in 2019, including total price, unit price, community name, and community location.

\item Chinese Major Cities' POI Data~\cite{poi_data}: The information on various types of POIs within China's 28 major cities was collected from Gaode Map in 2019, including store name, type, location, area, and street information.

\end{itemize}

\section{Statistical Analysis Based on Data}
\textbf{Identifying commuters from the same household.}
To obtain the actual number of commuter households from the data, we take reference of the 7-th Census of Shijiazhuang~\cite{hbpopcensus}, revealing the proportion of households from one-person to six-person to be 0.12, 0.27, 0.23, 0.19, 0.097, and 0.093 respectively. According to the demographic yearbook~\cite{census2017}, one commuter can support up to 1.92 people, hence 422,454 commuters can support a maximum of $N_{p} = 811112$ people. Denoting $p_i$ is the proportion of households with $i$ members, the number of households with $i$ members is $n_i=p_i\times N$, where  $N$ is the total number of households. The number one-person to six-person households are found to be 30,870, 69,458, 59,168, 48,878, 24,953, and 23,924 respectively.

Next, we assumed that the number of commuters in one household is just the supply minimum $n_c^{min}$, or $n_c^{min}-1$, without considering supply redundancy. Therefore, the potential number of commuters in  households with different number of members is shown in the Supplementary Table S1.

\begin{table}[h]
\centering
\begin{tabular}{|c|c|c|c|c|c|c|}
\hline
family size  & 1 person & 2 persons & 3 persons & 4 persons & 5 persons & 6 persons \\ \hline
0 commuter   & $\surd$     &       &       &       &       &       \\ \hline
1 commuter   & $\surd$     & $\surd$     & $\surd$     &       &       &       \\ \hline
2 commuters   &       & $\surd$     & $\surd$     & $\surd$    & $\surd$    &       \\ \hline
3 commuters   &       &       &       & $\surd$     & $\surd$     & $\surd$     \\ \hline
4 commuters   &       &       &       &       &      &   $\surd$    \\ \hline
\end{tabular}
\caption*{{\bf Supplementary Table S1.} The possible commuter families in different household types.}
\label{tab:my_label}
\end{table}

Thus, the number of households with different number of commuters can be calculated, taking two-person households as examples:
\begin{equation}
\frac{N_{p}^{2}}{N_{c}^{2}} = \frac{2n_2}{n_{2}^{1} + 2n_{2}^{2}} = 1.92
\end{equation}
\begin{equation}
n_{2}^{1} + n_{2}^{2} = n_2
\end{equation}

where $N_{p}^{2}$ is the total number of people in $2$-person households; $N_{c}^{2}$ is the total number of commuters in $2$-person households ; $n_{2}^{j}$ is the number of $j$-commuter families in $2$-person households. Thus, the actual number of different commuter households in the data can be calculated, with the number of households from single-commuter to four-commuter households to be 108,515, 90,843, 40,076, and 3,006 respectively.

\textbf{Commuter Households Obtained through Trajectories Similarity.}
 Since individual mobility trajectory data do not include the information about household membership, we used data from the 4 days in the 2 weekends to identify commuters from the same household. According to the census information, most of households in Shijiazhuang have less than $7$ members, and each working person supports up to 1.92 individuals. Therefore, we assume that each household has a maximum of 6 members and 4 commuters. We identify family members by calculating the similarity between the mobility trajectories among individuals who share the same residence location.
By analyzing the distribution of similarities $S_{ij}$ between individuals, those with $S_{ij} \ge 0.5$ are considered household members.  An individual might have similarities over 0.5 with multiple people; these similarities are sorted from the highest to the lowest, and only a maximum of  five  individuals with the highest similarity are considered to be potential household members. Ultimately, the ratio of single-, double-, triple- and quadruple-commuter households was found to be 65:26:6:2.

Compared to our census-deduced household ratios of 37:31:14:1, double and triple-commuter households are under-estimated, while single and quadruple-commuter households are over-estimated. To match the constructed and the deduced household ratio, we randomly selected 3,302 quadruple-commuter households. In each selected quadruple-commuter household, we use the three members with the highest similarity in each household to form a triple-commuter household, with the remaining person forming the single-commuter household, resulting in 192,777 single-, 76,533 double-, 21,529 triple-, and 3,006 quadruple-commuter households. In this case,  the number of single-commuter households still exceeded the inferred amounts, due to the criteria of $S_{ij} \ge 0.5$, which is a more stringent criteria for identifying household members, and thus leads to an overestimation of single-commuter households.

To identify households with a ratio in agreement with the census of Shijiazhuang, we release the stringent criteria  and form households among individuals from single-commuter households with $S_{ij}\ge 0.1$. We finally obtained a ratio of households matching the deduced ratio, such that single to quadruple-commuter households to 108,515, 90,843, 40,076, and 3,006 respectively.

\textbf{Using the shortest travel distance by motor vehicle.}
The mode of transport is highly related to the  commuting distance, we analyzed the distance for on-demand motor vehicles  the Didi in China in 2017 (as shown in Figure S1) to identify a threshold for urban residents to differentiate between short and long distances.
As we can see in Figure S1, the number of on-demand travel peaks at a distance of 2.5 km. Therefore, we assume that 2.5 km is the shortest distance that urban residents opt for motor vehicle travel in their daily commuting travel between their home and workplace. This finding is in alignment with the 2020 National Commuting Report of China, in the classification of non-motorized vs. motorized travel~\cite{Baidu2020}.

\begin{figure}[h!]
  \centering
  \includegraphics[width=12 cm]{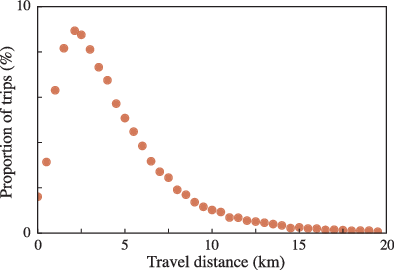}\\
   \textbf{Supplementary Figure S1.} The Distribution of Travel Distance of DiDi On-demand Vehicles in China. \label{FigS1}
\end{figure}

\textbf{Empirical commuter departure time.}
\begin{figure}[h!]
  \centering
  \includegraphics[width=16 cm]{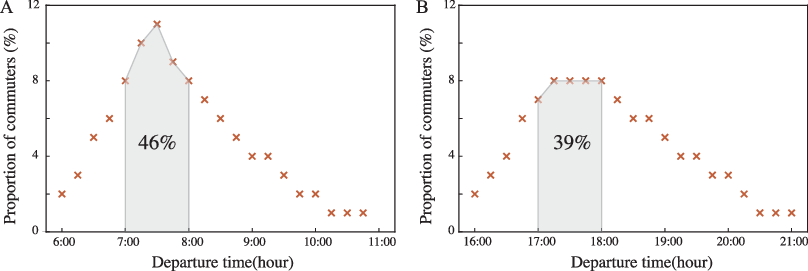}\\
   \textbf{Supplementary Figure S2.} The Distribution of Empirical Time of Departure  from Home and Workplace. Panel A shows the empirical time of departure of commuters from their home during the morning rush hour; Panel B shows the empirical time of departure from their workplace during the evening rush hour. \label{FigS2}
\end{figure}
Commuters' empirical departure time is set as the average time of their department from their home between 5 AM and 11 AM on the 5 weekdays in the dataset. The return time from the workplace for commuters is set as the average time of their departure from their workplace between 4 PM and 11 PM on the 5 weekdays in the dataset, as shown in Figure S2. Commuters departures are notably observed during the morning rush hour between 7-8 AM (Figure S2A), and most commuters returns are observed during the evening rush hour between 5-6 PM (Figure S2B), though the return time distribution is noticeably more dispersed.

\section{Greedy Home Swapping}

\textbf{Change in departure time.}
The empirical distribution of commuters' departure time is shown in Figure S2A. As we can see, the departure time of commuters peaks between 7-8 AM, with the highest number of departures at 7:30 AM. It is stipulated that if the morning arrival time to the workplace and the evening departure time from the workplace remain unchanged, the new morning departure time from their new home and the evening new arrival time back to their new home after GHS would shift and can be deduced from navigation data.

The distribution of the commuters' new departure time from their new home after GHS is shown in Figure S3A. The results revealed that the departure time becomes more dispersed compared to that in the empirical data, with fewer residents departing during peak time (7:30 AM). The distribution of the new departure time after SDGHS with medium tolerance is shown in Figure S3B, which is similar to that after GHS, but the number of commuters who use motor vehicles increases, resulting in a lower reduction in commuting distance, i.e. $\Delta d$.
\begin{figure}[h!]
  \centering
  \includegraphics[width=16 cm]{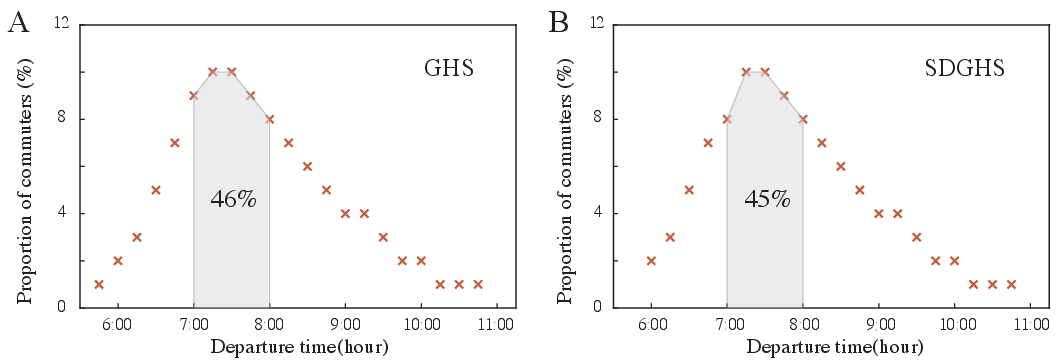}\\
   \textbf{Supplementary Figure S3.} The Distribution of Home-Departure Time During the Morning Rush Hour. Panel (A) shows the departure time after GHS; Panel (B) shows the departure time after SDGHS under medium tolerance.\label{FigS3}
\end{figure}

\textbf{Snapshots of congestion status obtained based on the empirical data and after RHS and GHS during rush hour.} By computing the congestion coefficient of each road $c_{\alpha}$ (see Materials and Methods in main text) during the morning peak hours between 7-8 AM and between 8-9 AM , and the evening peak hour between 5-6 PM obtained by the empirical data, and after RHS and GHS, we visualize the congestion coefficient in Figure S4. We found that the road congestion after RHS is greater than that based on the empirical data at all time, while the road congestion after GHS is less than that based on the empirical data. Moreover, after RHS, most road segments are congested, whereas after GHS, only a few segments remain congested, with most becoming slow-moving.
\begin{figure}[h!]
  \centering
  \includegraphics[width=16 cm]{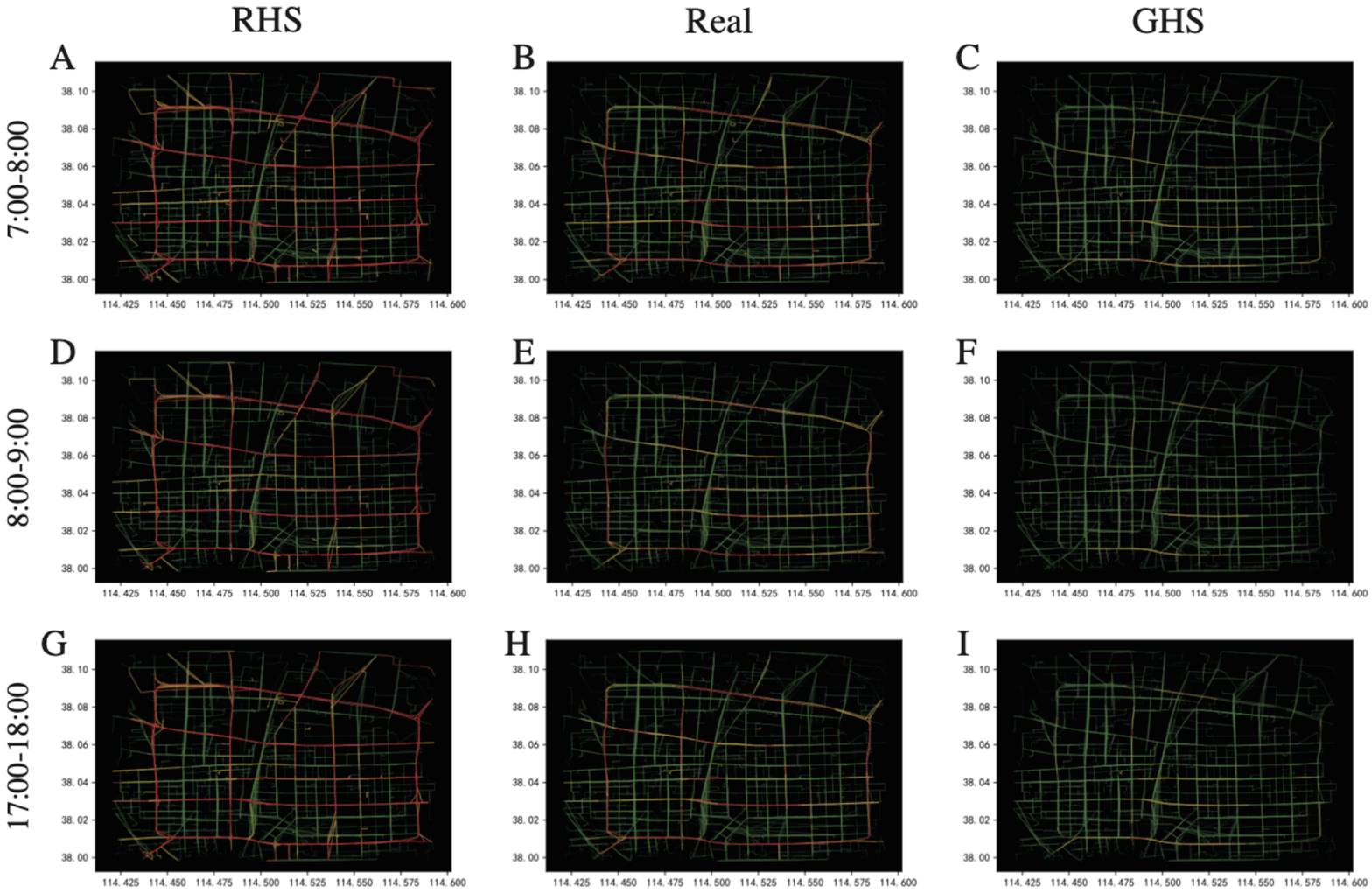}\\
   \textbf{Supplementary Figure S4.} The congestion coefficient obtained based on the empirical data and after RHS and GHS. The top three maps show the congestion coefficient during the morning peak hour between 7-8 AM; the middle three maps show the congestion coefficient during the morning peak hour between 8-9 AM; the bottom three maps show the congestion coefficient during the evening peak hour between 5-6 PM.\label{FigS4}
\end{figure}

\textbf{Average congestion coefficient of city affected by home swapping.}
\begin{figure}[h!]
  \centering
  \includegraphics[width=12 cm]{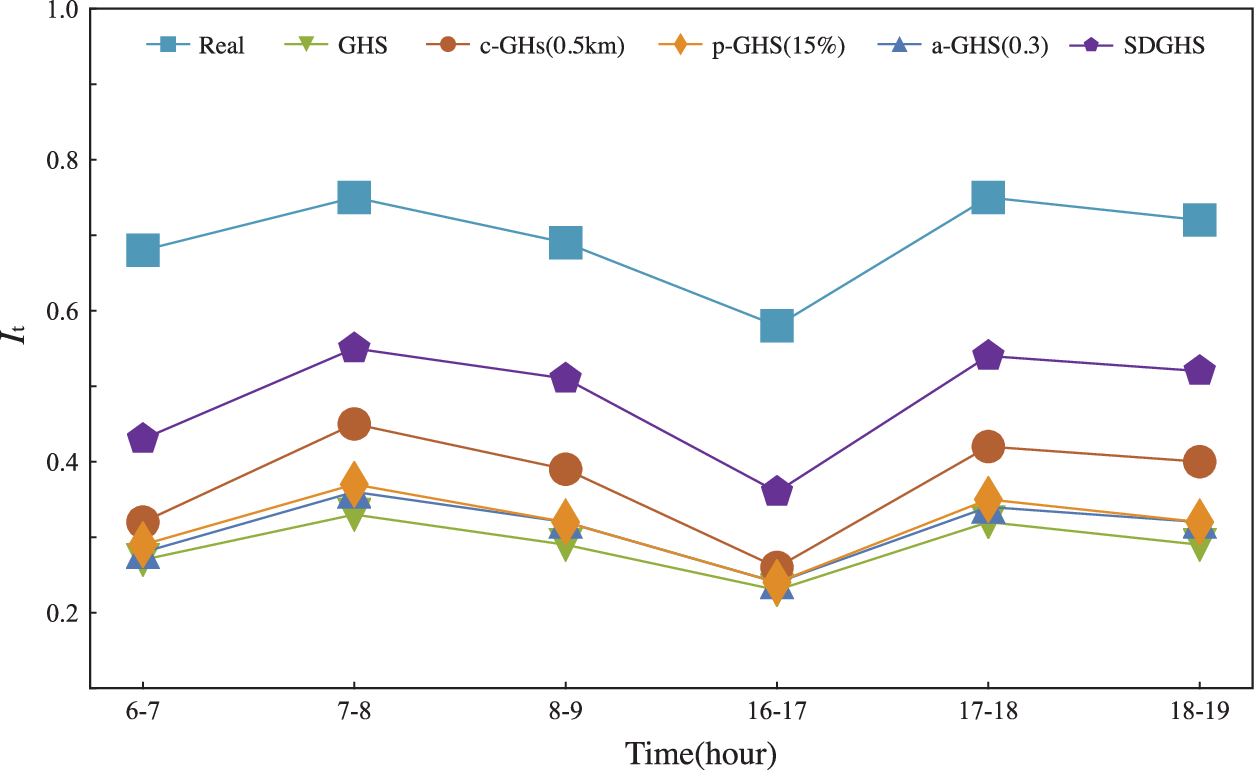}\\
   \textbf{Supplementary Figure S5.} The Average Congestion Coefficient of Shijiazhuang over Rush Hour.\label{FigS5}
\end{figure}
Empirical data show that the average congestion coefficient is highest during the morning rush hour (7-8 AM) and the evening rush hour (5-6 PM). The average congestion coefficient after RHS is significantly higher than that based on the empirical data; whereas $I_{t}$ after GHS is much lower than that based on the empirical data (see Figure 3C in the main text). $I_{t}$ is similar under medium tolerance on amenity accessibility and housing price, but after SDGHS under the medium tolerance, the city is more congested and closer to empirical data, as shown in Figure S5.

\begin{figure}[h!]
  \centering
  \includegraphics[width=16 cm]{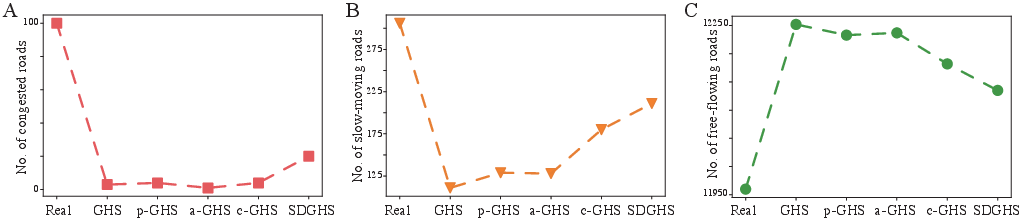}\\
   \textbf{Supplementary Figure S6.} The number of congested, slow-moving, and free-flowing road segments during the morning rush hour between 7-8 A.M. \label{FigS6}
\end{figure}
Since the average congestion coefficients $I_{t}$ after GHS under different factors do not vary significantly, we further analyze the changes in the number of congested roads with $c_{\alpha} \geq 1$, slow-moving roads with $0.4\leq c_{\alpha} <1$, and free-flowing roads with $c_{\alpha} < 0.4$ during the morning rush hour between 7-8 AM under different restrictive factors of GHS, as shown in Figure S6. After GHS, there is a noticeable decrease in congested and slow-moving road segments, while free-flowing road segments significantly increase. By considering only the amenity accessibility   (a-GHS), housing price (p-GHS) and the distance to the city center  (c-GHS), as well as all the three socio-demographic factors  (SDGHS), the average congestion coefficient $I_{t}$ gradually increases, the number of severely congested and slow-moving road segments gradually increases, while the number of free-flowing road segments decreases. Although the number of congested roads in a-GHS is lower than that after GHS, slow-moving segments increase, and free-flowing road segments decrease. Additionally, $I_{t}$ after p-GHS is similar to that after a-GHS, but with a higher number of congested and slow-moving segments.

\textbf{Considering dual socio-demographic factors  (medium tolerance) during home-swapping.}
 We also consider $\Delta d$ by considering two out of the three socio-demographic factors in GHS, as shown in Figure S7. In this discussion, we use the medium tolerance SDGHS discussed in Figure 4 of the main text. The reduction in average commuting distance, i.e. $\Delta d$ under (1) the joint consideration of the distance to the city center and amenity accessibility (c\&a-GHS) is 23.6\%, (2) the joint consideration of the distance to the city center and housing price (c\&p-GHS) is 26.6\%, and (3) the joint consideration of housing price and amenity accessibility (p\&a-GHS) reaches 36.29\%. The results show that the distance to the city center has a significant impact on the benefit brought by GHS, followed by amenity accessibility, and finally the housing price.

\begin{figure}[h!]
  \centering
  \includegraphics[width=12 cm]{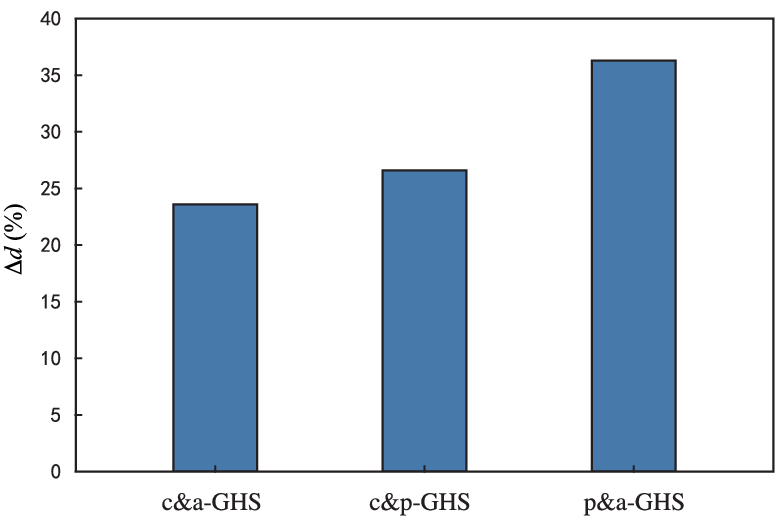}\\
   \textbf{Supplementary Figure S7.} The Reduction of the Average Commuting Distance, i.e. $\Delta d$, under the joint consideration of two out of the three socio-demographic consideration.\label{FigS7}
\end{figure}

\textbf{The impacts of the willingness and selfishness to swap home.} In GHS, we assume that all households are willing to swap their homes as long as the total commuting distance of the two swapping households decreases, even at the expense of an increase in the commuting distance of one of them. Here, we assume that the households with an increase of commuting distance has an $w\%$ willingness to swap but $1-w\%$ to reject.
\begin{figure}[h!]
  \centering
  \includegraphics[width=16 cm]{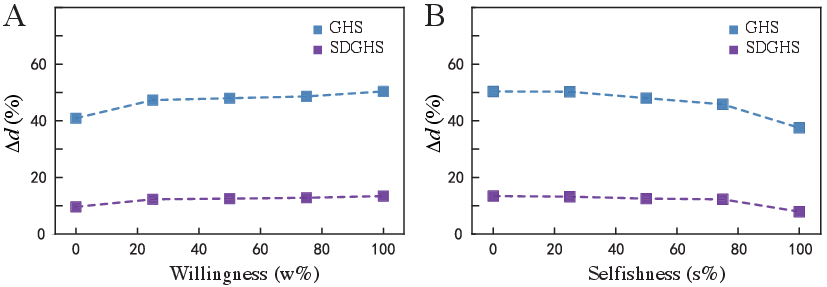}\\
   \textbf{Supplementary Figure S8.} (A) $\Delta d$ after GHS or after SDGHS (medium tolerance) considering households' willingness to change residences. (B) $\Delta d$ after GHS or after SDGHS (medium tolerance) considering household members' selfishness to change residences.\label{FigS8}
\end{figure}
As different households have different levels of willingness to swap homes, $\Delta d$ after GHS method with different willingness $w\%$ is shown by the blue line in Figure S8A. The results show that $\Delta d= 40.9\%$ when households do not swap residences if the average commuting distance of one household increases. As the willingness to swap increases from $0\%$ to 25\%, there is a noticeable increase in $\Delta d$. However, from a willingness of $25\%$ onwards, the increase in $\Delta d$ slows down as the willingness of swapping $w\%$ increases.

Moreover, we consider households willingness to swap home in SDGHS under medium tolerance and analyze $\Delta d$ after home swapping under different willingness $w\%$, as shown by the purple line in Figure S8A. The results find that as the $w\%$ increases, $\Delta d$ also increases. When $w=0$, $\Delta d$ is $9.6\%$; when $w>= 25$, there is no significant increase in $\Delta d$.

In GHS, we further consider the impact of household members' selfishness. It is assumed that a household member who experiences an increase in commuting distance will have $s\%$ selfishness to prevent swapping home, even if the average commuting distance of the household decreases. $\Delta d$ after GHS method with different selfishness $s\%$ is shown by the blue line in Figure S8B. The results indicate that $\Delta d$ = $37.5\%$ when households do not swap residence if the commuting distance of a household member increases. As the household members' selfishness increases, there is a decrease in $\Delta d$. We also consider the impact of household members' selfishness on SDGHS under medium tolerance and analyze $\Delta d$ after home swapping under different selfishness $s\%$, as shown by the purple line in Figure S8B. The results show that as the household members' selfishness increases, $\Delta d$ gradually decreases. When s = $100$, $\Delta d$ is $7.9\%$.

\section{The Impact of Household Special Needs to Home Swapping}
\textbf{The impacts of educational resources to home swapping.} In addition to considering the benefits to commuters, we also considered the impact of educational resources on home swapping. We randomly selected a proportion p of households who refused to relocate, regardless of commuting benefits, and swapped home only for the remaining household ($1-p$). Figure S9A shows the percentage decrease in the average commuting distance after home swapping. We find that, as the proportion p of households who refuses to relocate increases, $\Delta d$ decreases in both GHS and SDGHS (medium tolerance). The results in the figure are averaged over ten independent simulations, It is worth noting that even in the extreme scenario $p$ = $90$ (i.e.,the vast majority of households are immovable), $\Delta d$ still reaches $8.5\%$ in GHS and $2.3\%$ in SDGHS. This suggests that our framework remains effective in reducing congestion even when many households prioritize non-commuting factors. Furthermore, we find that the decrease in $\Delta d$ in SDGHS is significantly slower than that of GHS with increasing p. This phenomenon can be attributed to the pivotal role of "educational resources" in POI within the context of amenity accessibility in SDGHS. In other words, the results of the simulation can be used to demonstrate again that SDGHS can better portray the impact of the household's need for a living environment during the process of swapping homes.
\begin{figure}[h!]
	\centering
	\includegraphics[width=16 cm]{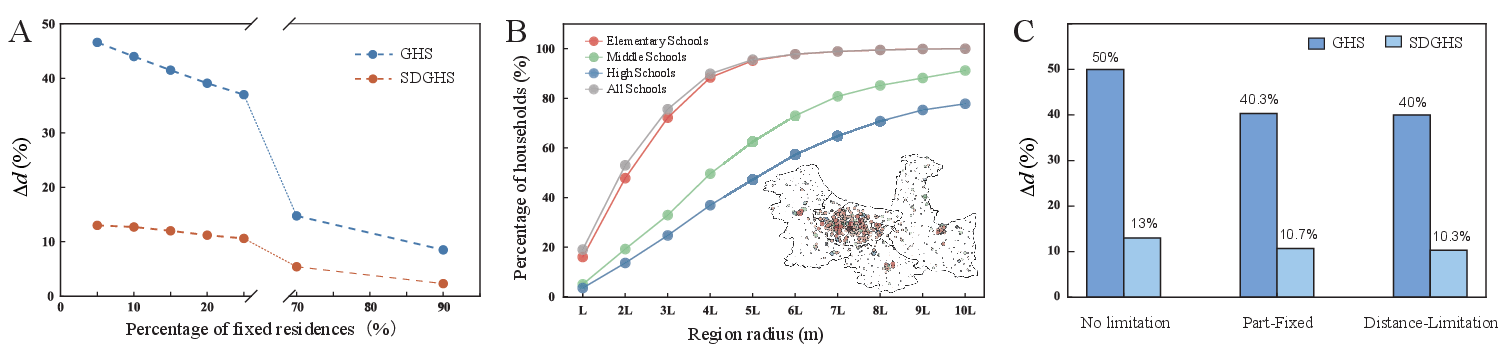}\\
	\textbf{Supplementary Figure S9.} Impact of residential constraints on commuting distance optimization. (A) Percentage reduction in commuting distance achieved through home-swapping optimization when randomly selecting a percentage of households to remain fixed in their residential locations. (B) Spatial distribution of schools and residential patterns: (top) proportion of families living within increasing region radii (L, 2L...10L) from elementary, middle, high, and all schools, respectively; (bottom) visualization of all schools where circle size scales with student population recorded during the study period. (C) The percentage decrease in average commuting distance under different educational resource constraints, comparing Greedy Home-Swapping (GHS) and Socio-demographic Greedy Home-Swapping (SDGHS) approaches: 'No limitation' (baseline without educational constraints), 'Part-Fixed' (fixing residences according to Hebei Province's 6-18 age population distribution), and 'Distance-Limitation' (binding households to remain within region radius $L$ of their nearest school). All results represent averages from 10 independent simulations, with SDGHS using medium tolerance parameters.\label{FigS9}
\end{figure}

To better align with real-word constraints, we analyzed the spatial distribution of educational facilities in the target city (Figure S9B). Using the geographic locations of elementary, middle, and high schools, we created concentric regions around each school with radii increasing by $N \times L$ (where L is the city's average street length) and calculated the proportion of households residing within these regions. The inset map visualizes the city's hierarchical distribution of educational resources. We then designated households within L meters of any school as "education-anchored" households, restricting them from participating in residential swaps. As shown in Figure S9C (Distance-Limitation scenario), this constraint affected $19\%$ of households when considering primary and secondary schools. Under this condition, GHS achieved a $40\%$ reduction in commuting distance, while SDGHS still yielded a $10.3\%$ optimization. In addition, according to Hebei Province's Seventh Population Census, $18\%$ of the population falls within the 6-18 age range (school-attending children). Assuming one student per eligible household, we randomly fixed $18\%$ of urban households to simulate education-driven immobility (Part-Fixed scenario). In this case, GHS reached a $40.3\%$ optimization rate, and SDGHS achieved $10.7\%$. The results presented in the figure have been obtained through the averaging process of 10 independent simulations.

\section{The Reduction of Carbon Emissions}

\begin{figure}[h!]
  \centering
  \includegraphics[width=16 cm]{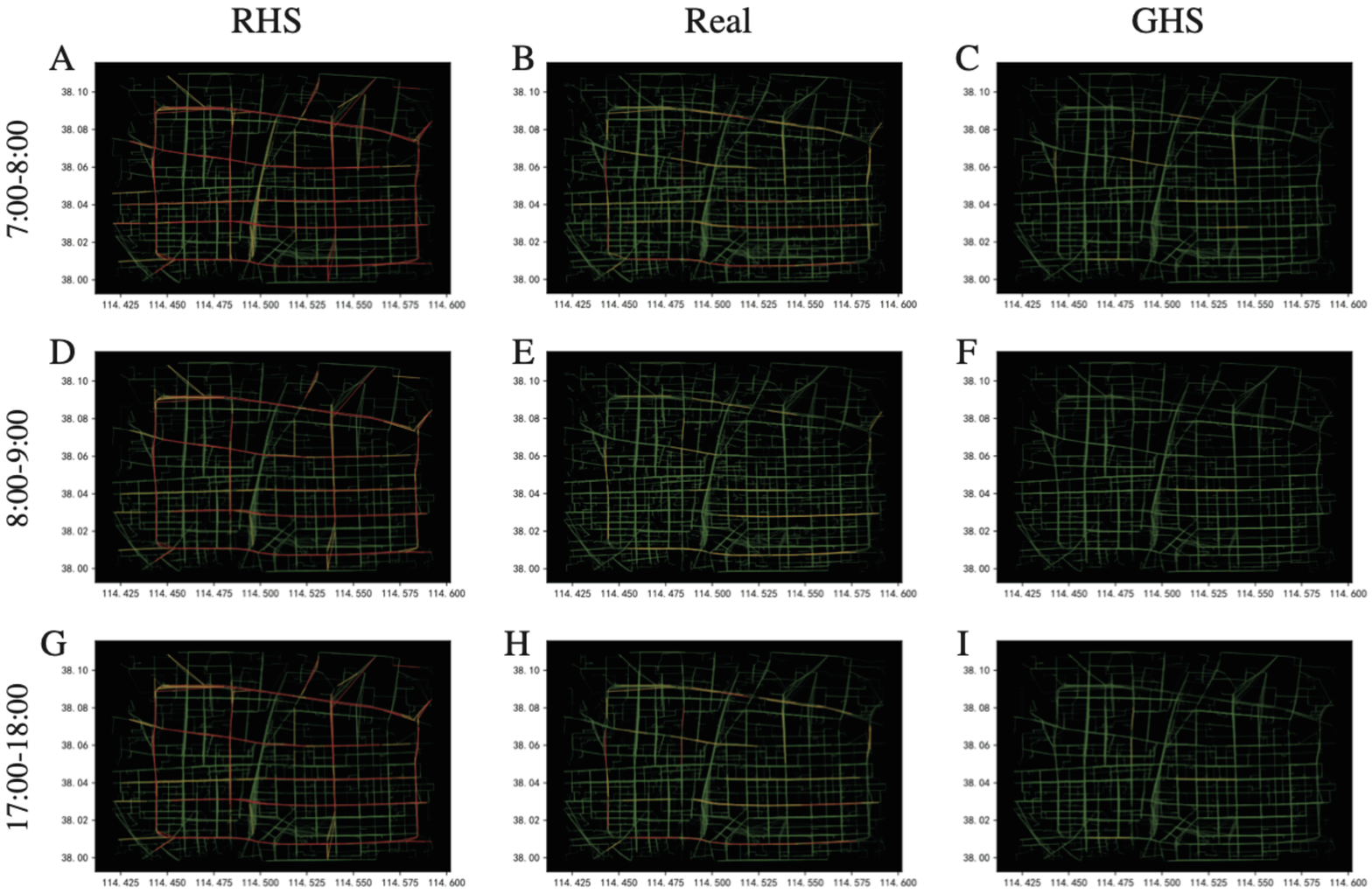}\\
   \textbf{Supplementary Figure S10.} The Road carbon emissions obtained based on the empirical data and after RHS and GHS. The top three diagrams show the CO$_2$ emissions on the roads during the morning peak hours between 7-8 AM; the middle three diagrams show the CO$_2$ emissions during the morning peak hours between 8-9 AM; the bottom three diagrams show the CO$_2$ emissions during the evening peak hours between 5-6 PM. \label{FigS10}
\end{figure}
We computed and visualized the carbon emissions of each road segment during the morning peak hours between 7-8 AM, between 8-9 AM, and the evening peak hour between 5-6 PM based on the empirical data and after RHS and GHS, as shown in Figure S10. We found that the carbon emissions under RHS are greater than that based on the empirical data at all time, while those after GHS are significantly less than that of the empirical data. The carbon emissions between 7-8 AM during the morning peak hours are higher than those between 5-6 PM during the evening peak hours. Additionally, after RHS, most road segments emit a large amount of CO$_2$, whereas after GHS, only a few segments have higher emissions, and most road segments emit a small amount of CO$_2$ (road segments emitting 0.4 tonnes of carbon are marked in red, and those emitting between 0.2 and 0.4 tonnes are in yellow, and those emitting less than 0.2 tonnes are in green).

\begin{figure}[h!]
  \centering
  \includegraphics[width=12 cm]{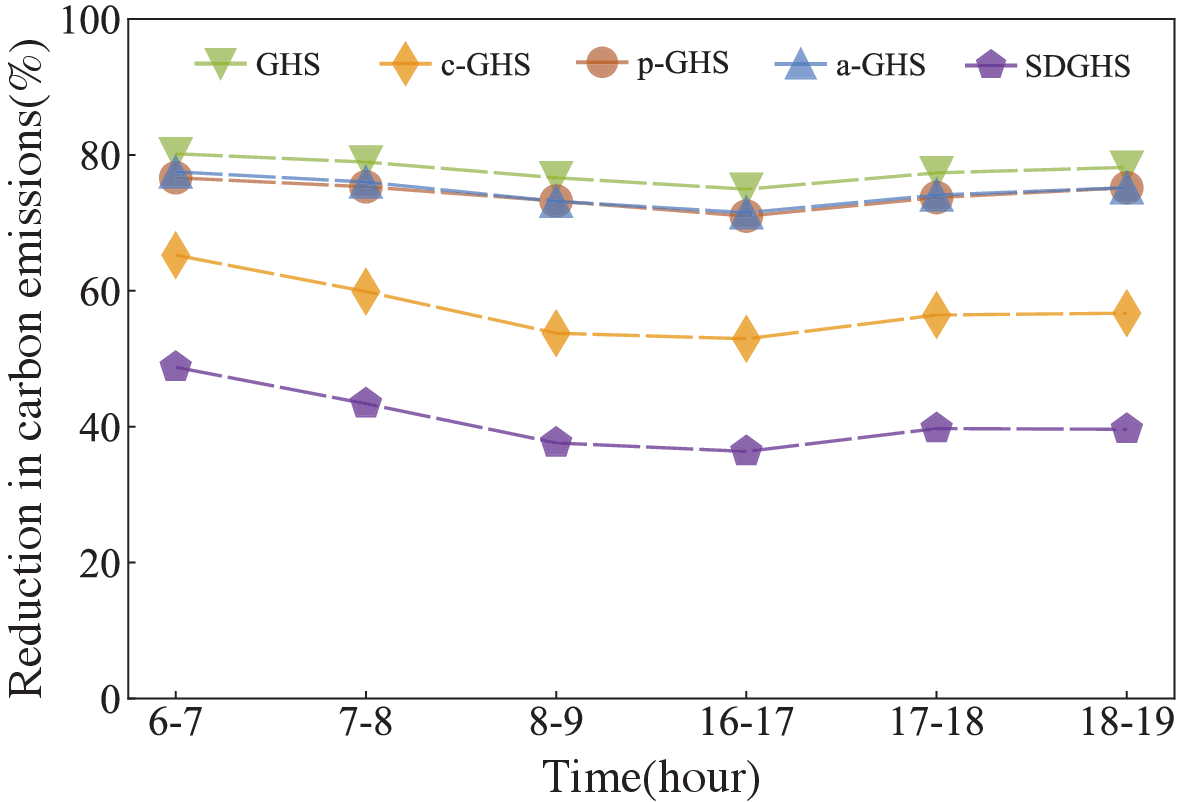}\\
   \textbf{Supplementary Figure S11.} The carbon emissions during rush hour under different conditions.\label{FigS11}
\end{figure}
The reduction in carbon emissions after GHS, and SDGHS is shown in Figure S11 for each hour during the peak commuting hours. In the morning peak hours, the average commuting distance after home swapping become shorter for most households. After home swapping, each individual is supposed to arrive at their workplace at the same time before home swapping, which leads to a dispersed departure time (Figure S3). It was found that the reduction in carbon emissions decreases over time. Similarly for the evening peak hours, where each individual is supposed to leave workplace at the same time after home swapping, the shorter commuting distances lead to a gradual increase in carbon emissions reduction over time. Among all the cases, the GHS emissions reduction is the highest, and the emissions reduction of SDGHS is the lowest.

\begin{figure}[h!]
  \centering
  \includegraphics[width=12 cm]{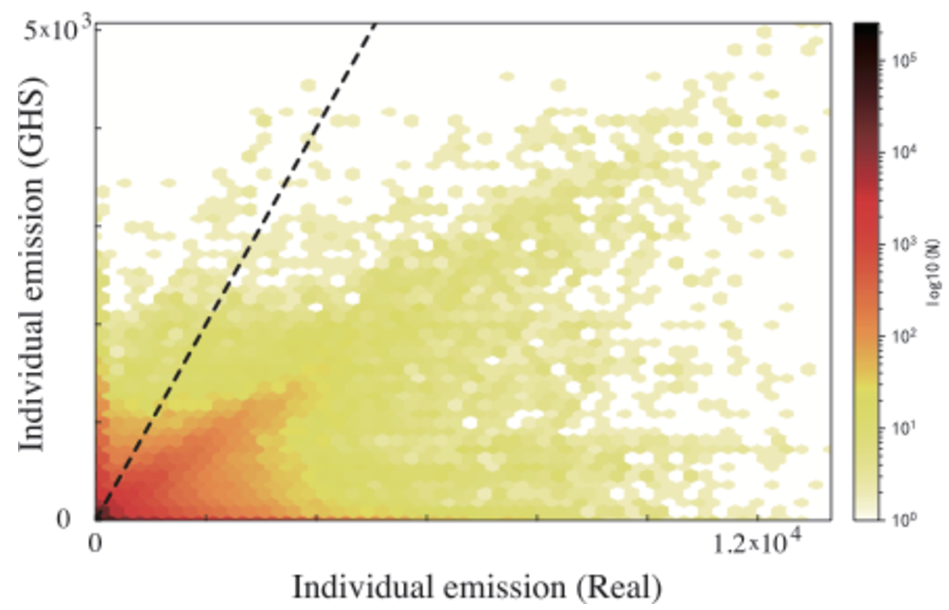}\\
   \textbf{Supplementary Figure S12.} The heatmap of changes in individual emissions  before and after GHS.\label{FigS12}
\end{figure}
We further analyze the changes in commuters' emissions before and after residential allocation by GHS, as shown in Figure S12. The individual's unaltered emissions following GHS is indicated by the black dashed line, and the majority of commuters experience a reduction in emissions after changing their residence, as changes in residence lead to non-motorized travel, reducing emissions to zero. However, there are cases where individuals with initially short commuting distances experience an increase in both commuting distance and emissions after changing residence.

\section{Adding New Activity Centers based on the Original City Center}

(1) Adding new activity centers and the corresponding amenity configurations. In this study, different numbers of new activity centers are established in surrounding county towns, and the amenity configurations around the new centers are changed accordingly. Based on the distance of each base station from the original city center and the proportions of various amenities, as represented by POI, around each base station, the corresponding POIs are added to the surrounding of the new centers, as shown in Figure S13. The yellow line corresponds to  an increase of the number of centers without the corresponding amenities, while the green line represents adding the corresponding POI configurations around the new activity centers. Comparing the two lines, there is not a significant increase in $\Delta d$ with the corresponding amenity configurations added.

\begin{figure}[h!]
  \centering
  \includegraphics[width=12 cm]{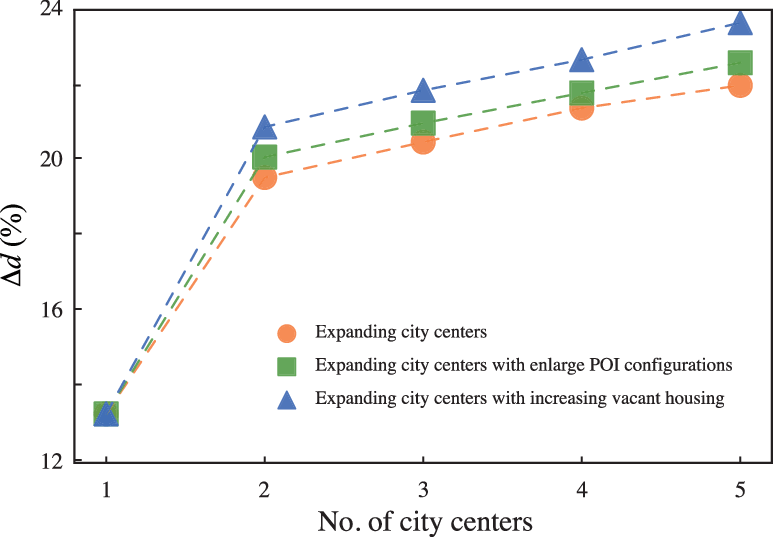}\\
   \textbf{Supplementary Figure S13.} $\Delta d$ after adding new activity centers, where the green line and symbols represent the addition of the corresponding amenity configurations, and the blue line and symbols represent the addition of the corresponding vacant housing. \label{FigS13}
\end{figure}

(2) Adding new centers and increasing vacant housing. Without providing new workplaces, the residential capacity of the new centers is proportionally increased as in the original city center. As shown by the blue line in Figure S13, as the number of centers increases, $\Delta d$ after applying medium tolerance SDGHS gradually increases, showing an improvement compared to just adding new centers.

\section{Impact of Commuting Mode Shift Threshold}
After GHS, there are changes in the spatial configuration of residences and workplaces as well as commuters' commuting distances, which also lead to the corresponding changes in commuters' modes of transportation. Through analyzing the travel distance of on-demand Didi vehicles in China, we assume 2.5 km to be the shortest distance for commuters to travel by motor vehicle. It is considered that a commuter who travels less than 2.5 km does not contribute to traffic congestion and carbon emissions, and instead uses non-motorized transportation. First, we studied the impact of GHS under different restrictive conditions on the changes in commuters' modes of transportation, as shown in Figure S14. The results show that after residence re-allocation by GHS, 64\% of commuters used non-motorized transportation, which is significantly higher than the empirical number of non-motorized commuters (36.5\%). Under medium tolerance with three combined factors, 44.1\% of commuters used non-motorized transportation, which is only a 7.6\% increase compared to the empirical data.

\begin{figure}[h!]
  \centering
  \includegraphics[width=12 cm]{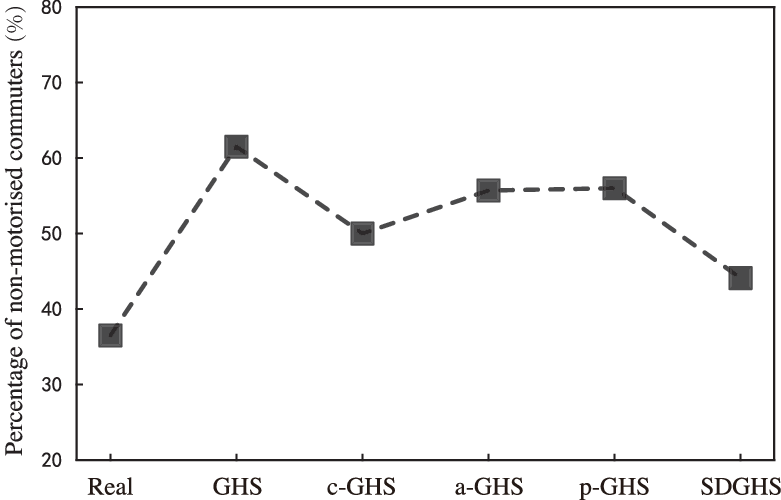}\\
    \textbf{Supplementary Figure S14.} The proportion of non-motorized vehicles used by commuters, after different versions of GHS under medium tolerance. \label{FigS14}
\end{figure}

While we have assumed 2.5 km to be the shortest distance for commuters to travel by motor vehicles, we also examined other thresholds of assumed minimum distance at  $2$, $1.5$, $1$, $0.5$, and $0$ km, as shown in Figure S15. As the threshold decreases, the proportion of non-motorized vehicles shows a significant decline. Under medium tolerance with SDGHS, 9\% of commuters live and work within a very short commuting distance.
\begin{figure}[h!]
  \centering
  \includegraphics[width=12 cm]{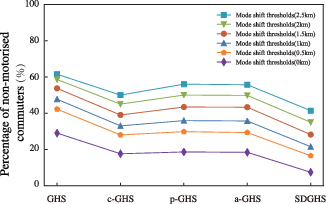}\\
    \textbf{Supplementary Figure S15.} The proportion of non-motorized vehicles used by commuters,  under different assumption thresholds of minimum distance in the use of motorized vehicle, after different version of GHS. \label{FigS15}
\end{figure}

In addition, we also explore the impact of the assumed thresholds of the minimum distance of the use of motorized vehicles on  the average congestion coefficient and carbon emissions. Figure S16A shows the impact of different values of thresholds on traffic congestion during the 7-8 AM peak hours. As the threshold value decrease, the number of used motor vehicles increases, the occupancy of road sections during peak hours increases, thus increasing congestion. When all commuters travel by motor vehicles, congestion also increases. By considering the three socio-demographic factors, there are fewer non-motorized vehicles, and the change in the minimum travel distance for using motor vehicles does not significantly impact traffic congestion.

\begin{figure}[h!]
  \centering
  \includegraphics[width=16 cm]{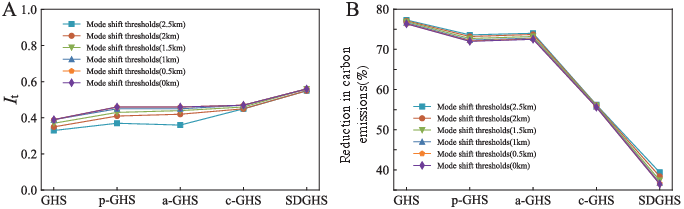}\\
   \textbf{Supplementary Figure S16.} The impact of the threshold values of the usage of the motorized vehicles. (A) shows the dependence of the average congestion coefficient $I_{t}$ on the threshold values. (B) shows the dependence of the reduction of carbon emissions on the threshold values. \label{FigS16}
\end{figure}

On the other hand, for carbon emissions, as the threshold decreases, the number of motor vehicles increases, and the carbon emissions increase, but not significantly, as shown in Figure S16B. It is because the reduction in the minimum travel distance for using motor vehicles increases the usage of motor vehicles commuting over a short distance, which does not significantly impact carbon emissions.

\section{The Impact of Induced Trips after Home Swapping}
Any traffic optimization leads to the emergence of new induced trips~\cite{law of road congestion,Time constraints,Spatio temporal,Flow improvements}. Therefore, we conducted preliminary analyses to examine how induced trips might impact our results. Using 15-minute discrete mobility trajectories, we identified potential induced travelers by analyzing both commuters who shifted their departure times ($8-10$ AM $\rightarrow$ $7-8$ AM) and non-commuters who adjusted their travel schedules. We simulated how congestion relief might generate new trips by randomly selecting $5\%-25\%$ of individuals who moved more than $2.5$ km and classifying them as motorized travelers, proportional to actual commuter volumes, and incorporating their real trajectory data.
Each synthetic traveler was assigned a morning departure time (7-8 AM) matching actual commuting patterns, with routes generated from their recorded stay points through navigation data. Through 10 repeated sampling iterations to ensure robustness, we found that while induced travel does gradually increase average congestion coefficients (Figure S17A), the values remain substantially lower than real-world baselines. The reduction in carbon emissions decreases to $71.7\%$ when considering $25\%$ induced trips (Figure S17B), though the system maintains significant benefits since average commuting distances (11.1km) far exceed induced trip lengths (7.6km). These results demonstrate that even accounting for temporal shifts in travel behavior, our home-swapping approach delivers meaningful improvements. We recognize that our current data cannot fully capture potential spatial induced trips (completely new routes enabled by reduced congestion), though the strong periodicity of weekday travel patterns suggests such effects would be limited. This comprehensive analysis confirms the robustness of our findings while transparently acknowledging the boundaries of our current modeling framework.

\begin{figure}[h!]
	\centering
	\includegraphics[width=16 cm]{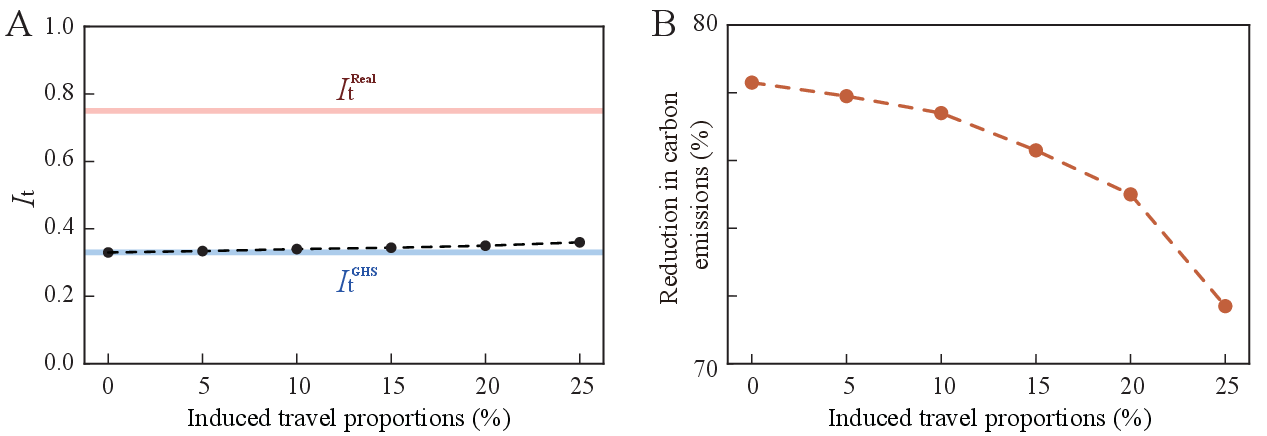}\\
	\textbf{Supplementary Figure S17.} Effects of induced morning travel on urban congestion and emission reduction. (A) City-wide average congestion coefficients under different induced travel proportions during 7-8 AM, where the induced travel proportion is defined as the ratio of newly added travelers to the original number of commuters. (B) Corresponding reduction in carbon emissions under varying induced travel proportions during the same morning peak period. All results represent averages from 10 independent simulation runs.\label{S17}
\end{figure}

\section{Impact of Multi-modal Transportation on Home Swapping}
We have addressed the over-simplification of assuming a single commuting mode by expanding our analysis to incorporate non-motorized transport, buses, and private cars, guided by empirical data from the 2017 Beijing Transport Development Annual Report~\cite{BeijingTransport2017}. Nevertheless, we remarked that the major city studied in the main text, i.e. Shijiazhuang, did not have a metro system at the time the dataset was obtained, and commuters were classified into three modes: trips under $2.5$ km as non-motorized, trips over $2.5$ km with an average speed $\leq 3$ m/s (consistent with reported bus speeds) as bus commuters, and trips over $2.5$ km with an average speed $>3$ m/s as private car users. This categorization aligned closely with official statistics ($18.5\%$ bus, $45\%$ car vs. reported $22.3\%$ bus, $37\%$ car), ensuring validity.

Based on the above analysis, the commuting mode of each commuter can be determined. If a commuter still travels a distance greater than $2.5$ km after the home swapping process, it is assumed that the commuter still maintains the original commuting mode. To model multi-modal impacts, we integrated the city's entire bus network - $246$ routes, $3661$ stops, and real-time schedules - into our simulations. This paper demonstrates that all bus routes within the urban area of Shijiazhuang were characterized using real-world stop trajectories and departure time data (Figure S18A). Bus trajectories during morning peak hours were reconstructed using navigation data, while car traffic flows were combined with bus traffic flows to compute congestion and emissions. It is important to note that the average length of a bus is four times that of a vehicle. Furthermore, given that the bus uses diesel, the carbon emission factor is 0.281 $kg CO_{2} e /km$.

\begin{figure}[h!]
	\centering
	\includegraphics[width=16 cm]{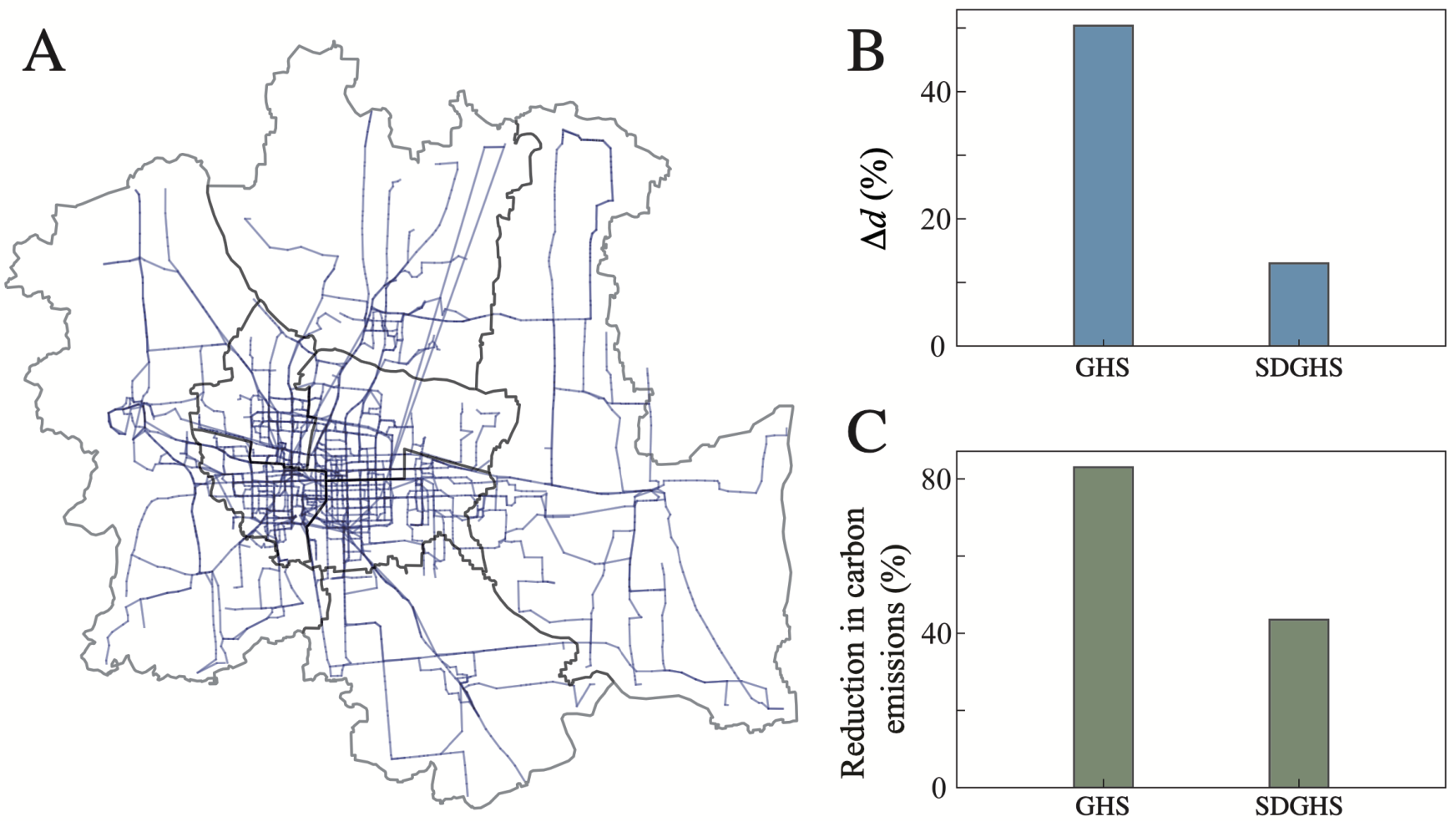}\\
\textbf{Supplementary Figure S18.} (A) Spatial distribution of all 246 bus routes (3,661 stops) in the Shijiazhuang, with route density indicated by color intensity (darker colors represent higher route density). Actual bus trajectories and schedules were incorporated from city transit data (see detail in SI section 1). (B) The reduction of average commuting distance by home-swapping (GHS and SDGHS) under multi-modal commuting conditions (medium tolerance). (C) The reduction of carbon emissions during morning peak hours under multi-modal conditions, with both GHS and SDGHS showing significant decreases in total city-wide carbon emissions despite accounting for bus and private vehicle interactions.\label{S18}
\end{figure}

Results in Figure S18B demonstrate that the core findings, with the assumption of multi-modal transportation, remain robust: GHS reduced the average commuting distances by ~$50\%$, while SDGHS (with medium tolerance for heterogeneity) achieved a $13\%$ reduction. Morning peak congestion decreased from 0.61 (empirical baseline) to $0.35$ (GHS) and $0.38$ (SDGHS), with corresponding declines in carbon emissions, $83\%$ (GHS) and $43.5\%$ (SDGHS), similar to that under the single-mode transportation (Figure S18C). These outcomes confirm that our proposed home-swapping approaches retain efficacy across diverse transportation modes.

\section{Commuting Surplus Analysis}
On the other hand, it is useful to incorporate a consumer surplus~\cite{Consumersurplus} analysis into our framework to quantify the economic impacts of home-swapping. We first define the fuel cost $T_i$ of an individual $i$ traveling through a set of paths $Path_{i}(t)$ is given by
\begin{equation}
	T_{i}=P_{gos} \times  {\textstyle \sum_{\alpha \in Path_{i}(t) }^{}d_{i}^{\alpha} \times f_{i}^{\alpha} }
\end{equation}
where $P_{gos}$ represents the fuel price ($7$ CNY per liter); $Path_i(t)$ is the set of road segments traversed by individual $i$ during time interval $t$; $d_{i}^{\alpha}$ is the length of road segment $\alpha$; $f_{i}^{\alpha}$ stands for the fuel consumption rate (liters per kilometer) of individual $i$ on road segment $\alpha$. Fuel consumption rates were parameterized according to roadway congestion levels: $0.165$ L/km (severely congested), $0.125$ L/km (moderately congested), and $0.085$  L/km (free-flow)~\cite{youhao}, as derived from real-time traffic coefficient data.

We then define the gain in daily commuting of individual $i$ after home-swapping to be
\begin{equation}
	T_{i}^{SL} = T_{i}^{Real} - T_{i}^{SDGHS}
\end{equation}
which is the difference between $T_i^{Real}$, the real daily commuting cost for individual $i$, and $T_i^{SDGHS}$, the daily commuting cost for individual $i$ after home-swapping under the SDGHS framework.

The gain $T_{i}^{SL}$ - defined as the difference between the commuting cost before and after home-swapping - reveals that over $80\%$ of households gain economic benefits under medium tolerance conditions (Figure S19), with monthly savings distributed broadly across income groups. This aligns with urban economics theory, as swaps can benefit individual households, even given our constraints to minimize housing cost disruptions (i.e. price differences $<15\%$) and to ensure most households retain comparable housing utility.

\begin{figure}[h!]
	\centering
	\includegraphics[width=16 cm]{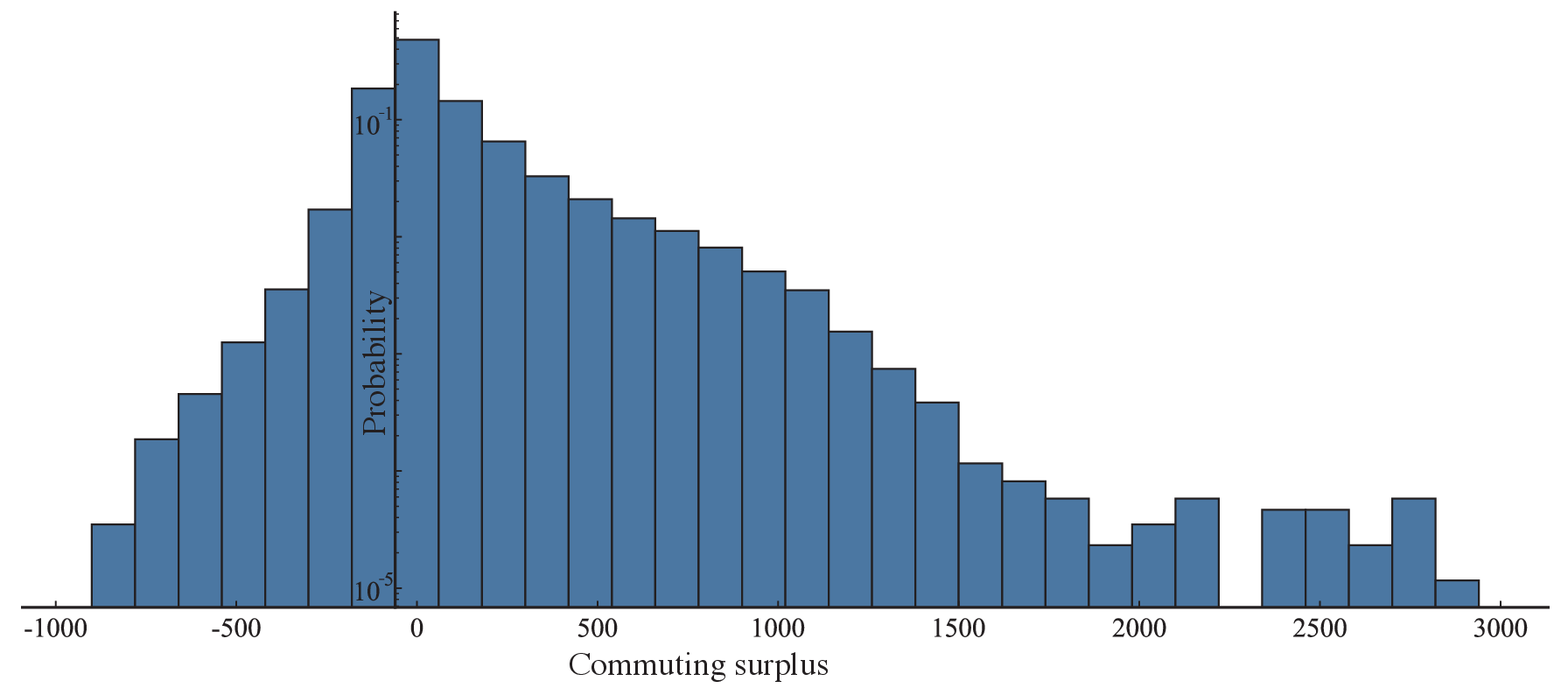}\\
	\textbf{Supplementary Figure S19.} Distribution of the commuting surplus $T_{i}^{SL}$ defined as the difference between original and post-swap commuting costs following the SDGHS optimization, with benefits sustained across income levels under medium ($15\%$) housing price change constraints. Over $80\%$ of households achieved a positive commuting surplus, demonstrating widespread economic benefits from the residential swapping process.\label{FigS19}
\end{figure}

\section{Examples of Home Swapping in the China's Major Cities}

Similar to Shijiazhuang, the initial analysis of the distributions of residence and workplace of commuters was conducted in the China's 28 major cities (see detail in Section 1). According to ref.~\cite{MajorCities_data2019}, the individual's residence is designated as the trip origin, and workplace is designated as the trip destination. The KL divergence between these distributions ranged from 0.06 to 0.18, with an average of 0.087 (Figure S20A), indicating that the distribution of residence and workplace is very similar across cities. Figure S20B presents heatmaps for Lanzhou, Nanjing, and Beijing. Similar to the pattern observed in Shijiazhuang, both residential and workplace locations are concentrated near the city center, while suburban areas show the opposite trend.

\begin{figure}[h!]
	\centering
	\includegraphics[width=16 cm]{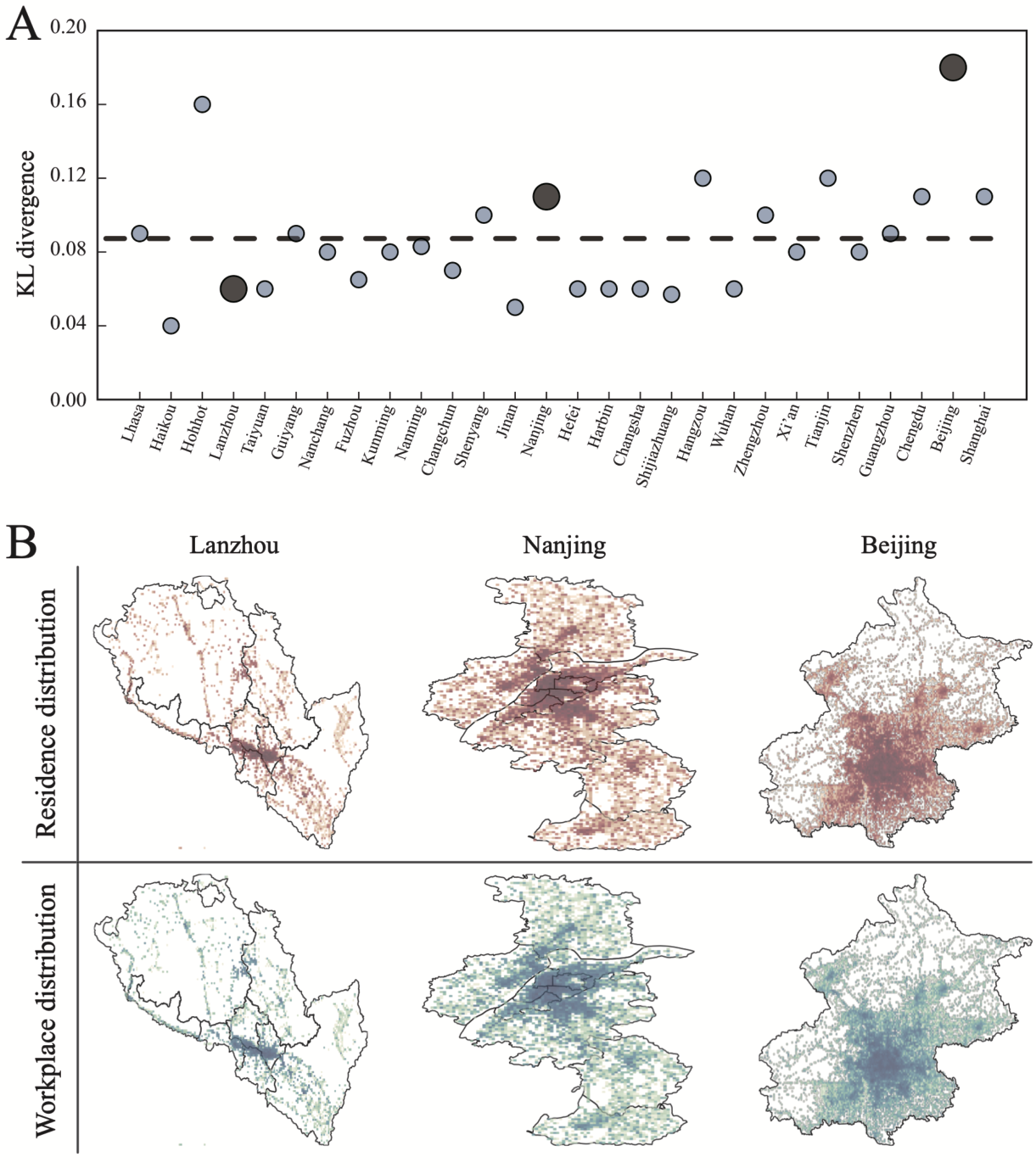}\\
	\textbf{Supplementary Figure S20.} Residential and Workplace Distribution in the China's 28 major cities. (A) KL divergence scatter plot of residential and workplace distribution. (B) The spatial distributions of residential and workplace in Lanzhou, Nanjing, and Beijing.\label{FigS20}
\end{figure}	
	
Due to the limitation of the origin-destination (OD) data without detailed individual trajectory information, it is impossible to infer household membership. Therefore, we constructed a realistic composite of 1-commuter and 2-commuter households based on the proportions reported in the 2019 China Statistical Yearbook~\cite{census2019}, as shown in Supplementary Table S2. Initially, commuters living at the same grid cell in the data were randomly paired to form 2-commuter households. A proportional random sample of these was subsequently extracted, with the remaining commuters as 1-commuter households. 
A decline in average commuting distance was observed after GHS across the 28 major cities of varying sizes (ranking by Seventh National Population Census~\cite{7NPC}). Notably, the correlation between the percentage decrease in average commuting distance ($\Delta d$) and city population size appears slightly positive. Interestingly, Chongqing showed an unusually high percentage decrease ($\Delta d = 88\%$) attributable to its unique mountainous topography, which suggests limitation on the applicability and accuracy of conventional planar geographic information systems for generating navigation. Due to the similar limitation of POI data and housing price data, we excluded Chongqing from the analysis of GHS with socio-demographic factors.

\begin{figure}[h!]
	\centering
	\includegraphics[width=16 cm]{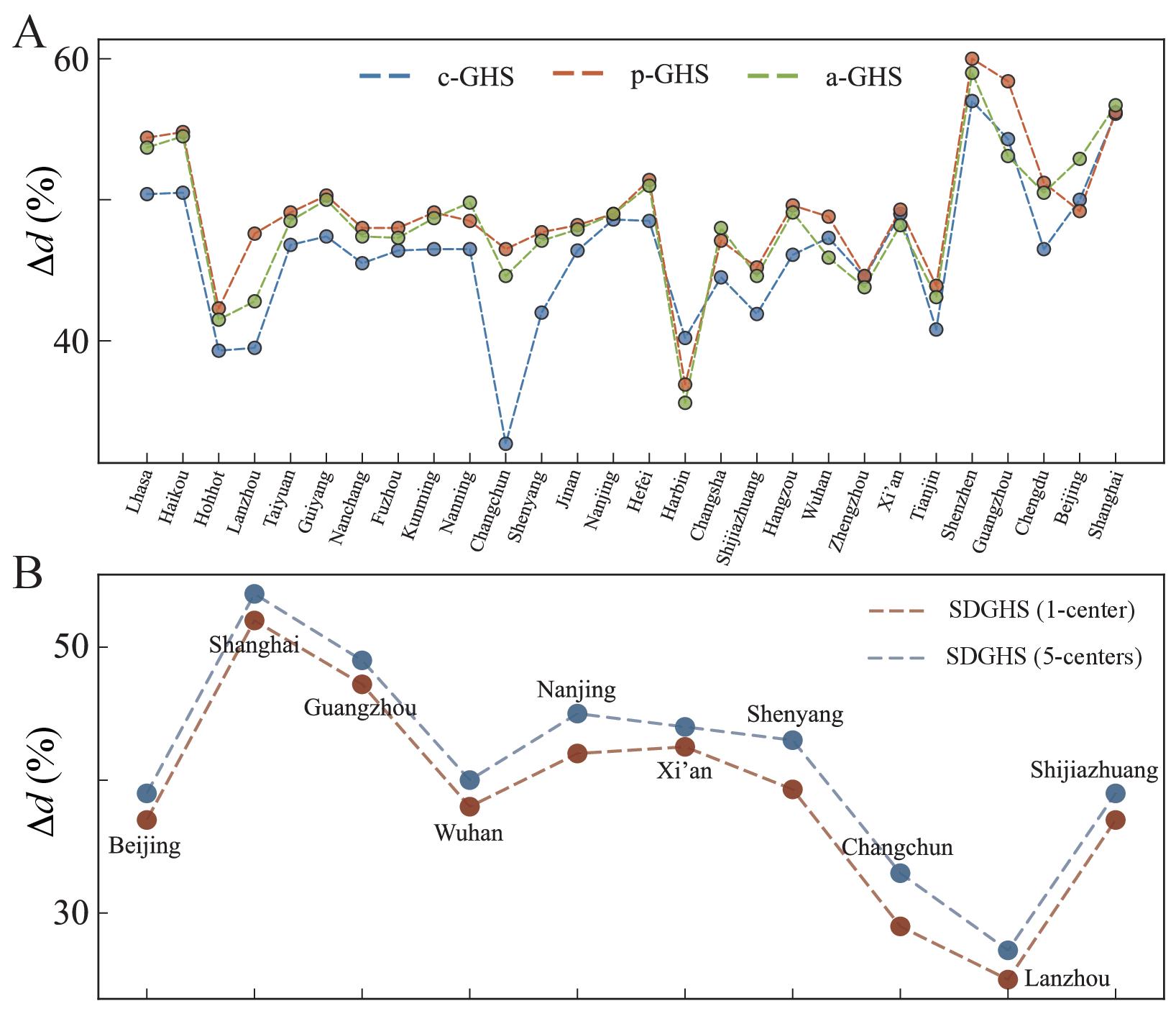}\\
	\textbf{Supplementary Figure S21.} GHS with Socio-demographic Consideration. (A) The percentage decrease in average commuting distance after c-GHS, p-GHS, and a-GHS in the China's 28 major cities. (B) The percentage decrease in average commuting distance after SDGHS and the impact of adding new centers in China's top-10 congested cities. All results represent averages from 10 independent simulations per city\label{FigS21}
\end{figure}

Then, we used data on amenities (POI), housing prices, and navigation routes of the 28 cities to analyze GHS considering a single socio-demographic. Given the $1$ km grid resolution, the constraints on distance to the city center, housing prices, and amenity accessibility were relaxed for household relocation. In addition to reducing commuting distance, two households swapped their homes only if: (1) The difference in their distance to city center (defined as the most visited grid cell between 7-8 AM) is less than 3 km (c-GHS); (2) The difference between their housing prices is less than $30\%$ (p-GHS); (3) The difference in amenity accessibility between the two residences is less than 0.3 (a-GHS). Amenity accessibility was calculated based solely on seven types of amenities within each grid cell, without considering neighboring cells. As shown in Figure S21A, the observed percentage decrease in average commuting distance after c-GHS, p-GHS, and a-GHS are detailed for the China's 28 major cities, which are ordered by population size.

As shown in Figure S21B, we also examine the SDGHS for China's top-10 congested cities, based on data from the Baidu Map Group's China Urban Transportation Report~\cite{congestion_cities}. In this analysis, Shijiazhuang replaces Chongqing due to the latter's special mountainous topography, as it is invalid to describe the regional features by using POI and housing price. To further investigate the impact of adding secondary centers, we identified the highest-traffic points in each administrative district and selected the top four locations outside the primary center as new centers. The results show that five-center cities achieved a greater reduction in average commuting distances after SDGHS than their single-center counterparts, consistent with the findings from Shijiazhuang.

Finally, we applied the same methodology to estimate the carbon emission reduction after GHS for the 10 most congested cities of China. Since the resolution of OD data is $1$ hour, it lacks the accurate individual departure time from his/her dwelling. To address this, we inferred departure time based on the OD matrix for the 7-8 AM period and actual commuting route navigation data. Specifically, if the navigated commute time exceeded one hour, the individual was assumed to depart at 7 AM. If it was less than one hour, a departure time was randomly assigned within the feasible window to ensure arrival by 8 AM.
Notably, when considering socio-demographic factors, the results of OD data show a greater improvement than those obtained by the high-resolution trajectories of Shijiazhuang in the main text. It can be attributed to several factors: Firstly, the coarse granularity of the data, constrained by a $1$ km $\times$ $1$ km grid resolution, homogenizes POI and housing price data within the grid, thereby reducing household heterogeneity to loosen the restrictions of home swapping. Secondly, due to the limited temporal resolution of the OD data, the individual departure time is estimated, not the real record, which underestimates the temporal crowdedness to flatten the peak-hour congestion. Thirdly, the current household matching mechanism is much more simplified and does not fully account for the complexity of real-world household structures, which lowered the difficulty of home swapping. Finally, the temporal coverage of the OD data may fail to capture cross-hour commuters outside the observation window.

\begin{table}[h]
\centering
\begin{tabular}{|c|c|c|c|c|}
\hline
\quad City \quad  & \quad $\#$grids \quad & \quad $\#$population (million) \quad & \quad Ratio ($\frac{\#2-\text{commuter}}{\#1-\text{commuter}}$) \quad \\ \hline
Shanghai & 2442 & 24.871 & 1.44\\ \hline
Beijing & 8717 & 21.893 & 1.37  \\ \hline
Chengdu & 5609 & 20.938 & 1.36    \\ \hline
Guangzhou & 3801 & 18.677 & 0.82  \\ \hline
Shenzhen & 1455 & 17.56 & 0.82  \\ \hline
Tianjin & 5671 & 13.866 & 1.99   \\  \hline
Xi'an & 2991 & 12.953 & 1.66  \\ \hline
Zhengzhou & 5507 & 12.601 & 1.82   \\  \hline
Wuhan & 4353 & 12.327 & 1.65   \\ \hline
Hangzhou & 5910 & 11.936 & 1.33  \\ \hline
Shijiazhuang & 3825 & 11.235 & 2.39   \\ \hline
Changsha & 5054 & 10.048 & 1.71   \\ \hline
Harbin & 2499 & 10.001 & 2.22  \\ \hline
Hefei & 1115 & 9.369 & 1.79  \\ \hline
Nanjing & 3880 & 9.315 & 1.82    \\ \hline
Jinan & 1052 & 9.202 & 2.14  \\ \hline
Shenyang & 1789 & 9.07 & 1.94  \\ \hline
Changchun & 2020 & 9.065 & 2.13  \\ \hline
Nanning & 5674 & 8.742 & 1.19   \\  \hline
Kunming & 3492 & 8.46 & 1.45  \\ \hline
Fuzhou & 7862 & 8.291 & 1.23   \\ \hline
Nanchang & 1733 & 6.255 & 2.03  \\ \hline
Guiyang & 1909 & 5.987 & 1.46   \\ \hline
Taiyuan & 1847 & 5.304 & 2.05  \\ \hline
Lanzhou & 2322 & 4.359 & 1.85   \\ \hline
Hohhot & 6152 & 3.446 & 2.39  \\ \hline
Haikou & 610 & 2.873 & 1.46   \\ \hline
Lhasa & 903 & 0.868 & 1.32  \\ \hline
\end{tabular}
\caption*{{\bf Supplementary Table S2.} The attributes of the China's 28 major cities.}
\label{tab:my_label}
\end{table}

\end{document}